\documentclass[11pt, draftclsnofoot, journal, letterpaper, onecolumn]{IEEEtran}

\usepackage[dvips]{graphicx}
\usepackage{times}
\usepackage{cite}
\usepackage{amsmath}
\usepackage{array}
\usepackage{amssymb}

\usepackage{stfloats}
\usepackage{rotating,threeparttable,booktabs}
\usepackage{bm}
\usepackage{dcolumn,booktabs}
\usepackage{xcolor}
\begin{document}

\title{Joint Source and Relay Precoding Designs for MIMO Two-Way Relaying Based on MSE Criterion}

\author{\IEEEauthorblockN{Rui~Wang and Meixia Tao*, \IEEEmembership{Senior Member,~IEEE}}
\thanks{Copyright (c) 2011 IEEE. Personal use of this material is permitted. However, permission to use this material for any other purposes must be obtained from the IEEE by sending a request to pubs-permissions@ieee.org.}
\thanks{The authors are with the Department of Electronic Engineering at Shanghai Jiao Tong University, Shanghai, 200240,
P. R. China. Emails:\{liouxingrui, mxtao\}@sjtu.edu.cn.}
\thanks{This work is supported by the NSF of China under grant 60902019, the Joint
Research Fund for Overseas Chinese, Hong Kong and Macao Young Scholars
under grant 61028001, and the Innovation Program of Shanghai Municipal
Education Commission under grant 11ZZ19.}}

\maketitle

\begin{abstract}
Properly designed precoders can significantly improve the spectral efficiency of multiple-input multiple-output (MIMO) relay systems. In this paper, we investigate joint source and relay precoding design based on the mean-square-error (MSE) criterion in
MIMO two-way relay systems, where two multi-antenna source nodes exchange information via a multi-antenna amplify-and-forward relay node. This problem is non-convex and its optimal solution remains unsolved.
Aiming to find an efficient way to solve the problem, we first decouple the primal problem into three tractable sub-problems, and then propose
an iterative precoding design algorithm based on alternating optimization.
The solution to each sub-problem is optimal and unique, thus the convergence of the iterative algorithm is guaranteed.
Secondly, we propose a structured precoding design to lower the computational complexity. The proposed precoding structure is able to parallelize the channels in the multiple access (MAC) phase and broadcast (BC) phase.
It thus reduces the precoding design to a simple power allocation problem.
Lastly, for the special case where only a single data stream is transmitted from each source node,
we present a source-antenna-selection (SAS) based precoding design algorithm.
This algorithm selects only one antenna for transmission from each source and thus requires lower signalling overhead.
Comprehensive simulation is conducted to evaluate the effectiveness of all the proposed precoding designs.
\end{abstract}


\section{Introduction}
Relay-assisted cooperative transmission can offer
significant benefits including throughput enhancement, coverage extension and power reduction in wireless communications.
It is therefore considered as
a promising technique for the next generation wireless communication systems,
such as LTE-Advanced and WiMAX.  Depending on whether the relay can receive and forward signals at the same time and frequency, there are two relay modes: full-duplex mode and half-duplex mode.
Although the half-duplex relay is
more favorable for practical implementation,
it is
less spectrally efficient than full-duplex ones. For instance, it will take four time slots for two source nodes to exchange information
with the help of a half-duplex relay when
there is no direct link.
To overcome the spectral efficiency loss caused by the half-duplex constraint,
two-way relaying has been recently proposed \cite{Larsson2006,ZhangS2006,Rankov2007,Katti2008}. The notion of two-way relaying is to apply the principle of network coding at the relay so as to mix the signals received from two links for subsequent forwarding, and then apply at each destination the self-interference cancelation to extract the desired information.
In contrast to the conventional one-way relaying,
two-way relaying only needs two time slots to complete one round of information exchange. Two-way relay strategies can be broadly divided into two categories,
decode-and-forward (DF) and amplify-and-forward (AF), similar to those in one-way relaying. In DF-based two-way relaying,
the relay decodes each individual received bit sequence, combines them together using XOR or superposition coding for example
and then broadcasts to the two destinations.
Decoding directly the combined bits may further improve the performance.
In AF-based two-way relaying,
the relay simply amplifies the received superimposed signals and forwards to the destinations. Compared with the DF relay strategy, the AF relay strategy is more attractive for its simplicity of implementation.

The multiple-input multiple-output (MIMO) technique is
a significant technical breakthrough in wireless communications.
By employing multiple antennas at the transmitter or the receiver, one can significantly improve the transmission reliability by leveraging spatial diversity.
If multiple antennas are applied at both the transmitter and receiver sides,
the channel capacity can be enhanced linearly with the minimum number of transmit and receive antennas.
Among various MIMO techniques, transmit precoding
is able to exploit the spatial multiplexing gain efficiently in both single-user and multi-user communication systems by making
use of channel state information (CSI) at the transmitter.
Incorporating the MIMO technique into
two-way relaying
is expected to
further increase the system throughput.
To fully realize the benefits of MIMO and two-way relaying, efficient
transmit precoding by taking relay nodes into account is crucial.
In this paper, we consider joint design of source and relay precoding in the MIMO two-way relay system where each node is equipped with multiple antennas.

Recently, a few studies have focused on MIMO two-way relaying.
The first category is based on the DF relay strategy.
For example, in \cite{Hammerstrom2007}, the authors investigate and compare the capacity gain for two different re-encoding operations.
In \cite{Oechtering2009}, the boundary of capacity region of Gaussian MIMO two-way relay broadcast channels is derived.
Furthermore,
the authors in \cite{Esli2008a, Esli2008} extend the DF-based MIMO two-way relay protocol to multi-user and cellular networks.
From the aforementioned works, it is easy to find that the precoding design for MIMO two-way relaying under the DF relay strategy does not differ much from the conventional multi-user MIMO precoding and hence many existing techniques can be applied.
The second category is based on the AF relay strategy.
The authors in \cite{Zhang2009} develop an algorithm to compute the globally optimal relay beamforming matrix
for a system where only the relay node is equipped with multiple antennas
and characterize the system capacity region. In \cite{Li2011}, the optimal relay beamforming matrix is designed to minimize the total mean-square-error (MSE) of two sources. Under the same design criterion, the authors in \cite{Li2010} consider the scenario
with multiple multi-antenna relay nodes.
Different from \cite{Zhang2009,Li2011,Li2010}, the works
\cite{Lee2009,Xu2011,Timo2008}
consider a system where the two source nodes are also equipped with multiple antennas.
In \cite{Lee2009}, applying the gradient descent algorithm, an iterative scheme is introduced to find the suboptimal relay precoder for sum-rate maximization.
In \cite{Xu2011}, the authors consider joint source and relay precoding design to maximize the sum-rate.
In \cite{Timo2008}, the authors propose a relay transceive precoding scheme by using zero-forcing (ZF) and minimum mean-square-error (MMSE) criteria with certain antenna configurations.
The precoding of MIMO two-way relaying with AF strategy has also been extended to multi-user networks.
For example,
the authors investigate the optimal relay precoding design for a MIMO two-way relay system with multiple pairs of users in \cite{Joung2010}
and further study the user scheduling problem in \cite{Joung2010a}.
In \cite{Ding2011}, the authors design a new network-coded transmission protocol for the same model as \cite{Joung2010} by
combining ZF beamforming and signal alignment such that the intra-pair interference and inter-pair interference can be completely canceled.
Other than using multiple antennas on each node, another way to achieve spatial diversity for AF relay strategy is to employ network beamforming among multiple single-antenna relay nodes as in\cite{Shahbazpanahi2010,Schad2011,Havary-Nassab2010,Zeng2011,Shahbazpanahi2010a,Vaze2011}.
Nevertheless, the precoding design for AF MIMO two-way relaying is much more challenging than that for the DF case.

In this study, we focus on the joint precoding design at both the source and relay nodes for MIMO two-way relaying with AF strategy.
Our goal is to minimize the total mean-square-error
(Total-MSE) of two users by assuming linear processing at both the transmitters and receivers.
Different from \cite{Li2011,Li2010},
we consider a two-way relay system where both the source and relay nodes are equipped with multiple antennas.
Furthermore, we study the joint source and relay precoding design rather than relay precoding design only.
The main contributions of this work are as follows:

\begin{itemize}
\item Iterative precoding design:  The joint optimization of source and relay precoding for Total-MSE minimization is shown to be non-convex
and the optimal solution is not easily tractable.
We propose an iterative algorithm to decouple
the joint design problem into three sub-problems
and solve each of them in an alternating manner. In particular, we derive the optimal relay precoder in closed-form when source precoders and decoders are fixed.
Since each sub-problem can be solved optimally, the convergence
of the iterative algorithm is
guaranteed.

\item Channel-parallelization based precoding design:
We further propose
a heuristic channel parallelization (CP) based precoding design algorithm for certain antenna configurations.
This method applies two joint matrix decomposition techniques so as
to parallelize the channels in the multiple access (MAC) phase and broadcast (BC) phase, respectively, of two-way relay systems.
Certain structures are hence imposed on the source and relay precoders.
Based on the proposed structure, the joint precoding design is reduced to a simple joint source and relay power allocation problem.

\item Source-antenna-selection based precoding design for single-data-stream transimssion:
For the special case where only a single data stream is transmitted from each source,
we introduce a source-antenna-selection (SAS) based precoding design algorithm.
We find that the SAS based precoding design can even outperform the iterative precoding design
in certain scenarios and yet has lower signalling overhead.
\end{itemize}

The rest of paper is organized as follows. In Section II, the MIMO two-way relaying model is introduced.
The iterative precoding design algorithm is presented in Section III.
Section IV describes the channel parallelization method and corresponding power allocation algorithm.
The source-antenna-selection based precoding algorithm
for single data stream is
included in Section V. Extensive
simulation results are illustrated in Section VI. Finally, Section
VII offers some concluding remarks.

\emph{Notations}: Scalar is denoted by lower-case letter, bold-face lower-case letter is used for vector, and bold-face upper-case letter is for matrix. $\cal E[\cdot]$ denotes expectation over the
random variables within the bracket. $\otimes$ denotes the Kronecker operator. $vec(\cdot)$ and $mat(\cdot)$ signify the matrix vectorization operator and the corresponding inverse operation, respectively.
${\rm Tr}({\bf A})$, ${\bf A}^{-1}$ and ${\rm Rank}(\bf A)$ stand for the trace, the inverse and the rank of matrix ${\bf A}$, respectively, and ${\rm Diag}(\bf a)$ denotes a diagonal matrix with ${\bf a}$ being its diagonal entries. Superscripts $(\cdot)^T$, $(\cdot)^{*}$ and $(\cdot)^H$ denote transpose, conjugate and conjugate transpose, respectively.
${\bf 0}_{N\times M}$ implies the
$N\times M$ zero matrix and ${\bf I}_N$ denotes the $N \times N$ identity matrix.
$||{\bf x}||^2_2$ denotes the squared Euclidean norm of a complex vector ${\bf x}$. $|z|$ implies the norm of the complex number $z$, ${\Re}(z)$ and $\Im (z)$ denote its real and image part, respectively. ${\mathbb C}^{x \times y}$ denotes the space of $x \times y$ matrices with complex entries. The distribution of a circular symmetric complex Gaussian vector with mean vector $\bf x$ and covariance matrix ${\bf \Sigma} $ is denoted by ${\cal CN}({\bf x},{\bf \Sigma})$.

\section{System Model}
\begin{figure}[tbhp]
\begin{centering}
\includegraphics[scale=0.50]{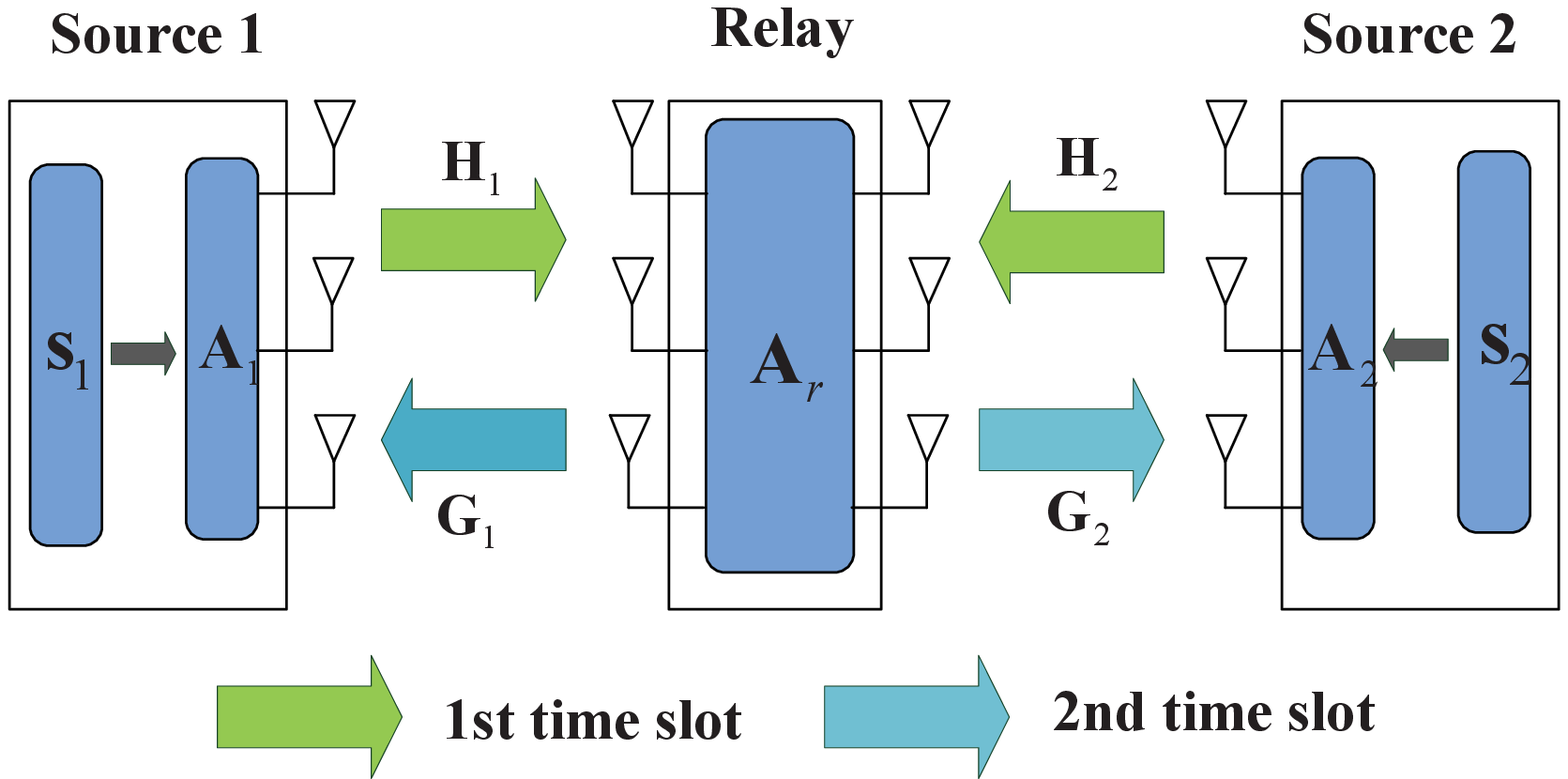}
\vspace{-0.1cm}
\caption{Illustration of the MIMO two-way relay system.} \label{fig:Two_way}
\end{centering}
\vspace{-0.3cm}
\end{figure}

Consider an $(N, M, N)$ MIMO two-way relay system where two source nodes, denoted as $S_1$ and $S_2$ and each equipped with $N$ antennas, want to exchange messages through a relay node, denoted as $R$ and equipped with $M$ antennas.
The information exchange
takes two time slots as shown in Fig.~\ref{fig:Two_way}.
In the first time slot (also referred to as the MAC phase), the two source nodes $S_1$ and $S_2$ simultaneously transmit the signals to the relay node $R$.
After receiving the superimposed signal,
the relay performs a linear processing by multiplying it with a precoding matrix and then forwards it in the second time slot (also referred to as the BC phase).
Without loss of generality, we assume that $N$ data streams are transmitted from each source in order to fully utilize the multiplexing gain. The special case with single data stream transmission shall be investigated in Section V.

Let ${\bf x}_i \in {\mathbb C}^{N\times 1}$ denote the transmit signal vector from source $S_i$, for $i=1, 2$. It can be expressed as
\begin{equation} \label{eqn:relayreceivedsignal}\nonumber
    {\bf x}_i={\bf A}_i{\bf s}_i,~i=1,2
\end{equation}
where ${\bf s}_i \in {\mathbb C}^{N\times 1}$ represents the information signal vector
with normalized power, i.e., ${\cal E}({\bf s}_i {\bf s}_i ^H)={\bf I}_{N}$,
and ${\bf A}_i \in {\mathbb C}^{N\times N}$ denotes the transmit precoding matrix.
Each column of ${\bf A}_i$ can be interpreted as the beamforming vector corresponding to the respective data stream in ${\bf s}_i$. The maximum transmission power at $S_i$ is assumed to be $\tau_i$, and thus we have
\begin{equation} \label{eqn:Powersource1}
{\rm Tr}\left({\bf A}_i {\bf A}_i^H\right) \leq\tau_i,~i=1,2.
\end{equation}

Let ${\bf y}_r$ denote the received $M\times1$ signal vector at the relay node during the MAC phase. It can be expressed as
\begin{equation} \label{eqn:relayreceivedsignal}\nonumber
    {{\bf y}_r}={\bf H}_1 {\bf x}_1 + {\bf H}_2 {\bf x}_2 + {\bf n}_r,
\end{equation}
where ${\bf H}_i \in {\mathbb C}^{M \times N}$ is the full-rank MIMO channel matrix from $S_i$ to $R$, and ${\bf n}_r$ denotes the additive noise vector at the relay node, following the distribution ${\bf n}_r \sim {\cal CN}({\bf 0},\sigma^2_r{\bf I}_{M})$.

Upon receiving ${\bf y}_r$, the relay amplifies it by multiplying it with a precoding matrix $ {\bf A}_r \in {\mathbb C}^{M \times M}$. Therefore, the $M \times 1$ transmit signal vector from the relay node can be expressed as
\begin{equation} \label{eqn:relaytransmittedsignal}\nonumber
    {\bf x}_r={\bf A}_r  {\bf y}_r.
\end{equation}
The maximum transmission power at the relay node is assumed to be $\tau_r$, which yields
\begin{equation} \label{eqn:RelayPowerCons}
    {\rm Tr} \left\{ {\bf A}_r \left( \sum^2_{i=1} {\bf H}_i {\bf A}_i {\bf A}_i^H {\bf H}_i^H +
    \sigma^2_r {\bf I}_{M} \right) {\bf A}^H_r \right\} \leq \tau_r.
\end{equation}
Then the received signal at
$S_i$ during the BC phase can be written as
\begin{equation} \label{eqn:source1receivedsignalnew}
    {\tilde {\bf y}_i} ={\bf G}_i {\bf x}_r + {\bf n}_i
    ={\bf G}_i {\bf A}_r {\bf H}_i {\bf A}_i {\bf s}_i + {\bf G}_i {\bf A}_r {\bf H}_{\bar i} {\bf A}_{\bar i} {\bf s}_{\bar i}
     + {\bf G}_i {\bf A}_r {\bf n}_r + {\bf n}_i,~i=1,2
\end{equation}
where ${\bar i}=2$ if $i=1$ and ${\bar i}=1$ if $i=2$, ${\bf G}_i \in {\mathbb C}^{N\times M}$ is the full-rank channel matrix from $R$ to $S_i$, ${\bf n}_i$ denotes the additive noise vector at $S_i$ with ${\bf n}_i \sim {\cal CN}({\bf 0},\sigma^2_i{\bf I}_{N})$.
Subtracting the back propagated self-interference term ${\bf G}_i {\bf A}_r {\bf H}_i {\bf A}_i {\bf s}_i$ from
\eqref{eqn:source1receivedsignalnew}
yields
the equivalent received signal vector at each destination node as
\begin{equation} \label{eqn:source1receivedsignalnewnew}
    {\bf y}_i = {\bf F}_i  {\bf s}_{\bar i}
     + {\bf G}_i {\bf A}_r {\bf n}_r + {\bf n}_i,~i=1,2
\end{equation}
where ${\bf F}_i= {\bf G}_i {\bf A}_r {\bf H}_{\bar i}{\bf A}_{\bar i}$ is the equivalent end-to-end MIMO channel matrix for $S_i$.

The problem in this study is
joint design of the precoding matrices $\{{\bf A}_1, {\bf A}_2, {\bf A}_r\}$ given the global CSI $\{{\bf H}_1, {\bf H}_2, {\bf G}_1, {\bf G}_2\}$
based on the MSE criterion. Specifically,
the objective is to minimize the Total-MSE of all the data streams of two users.
The Total-MSE has been widely chosen as a criterion for precoding design in the literature, e.g., \cite{Timo2008, Joung2010a, Joung2010, RonghongMo2009, WeiGuan2008, Tseng2009, Hunger2009}. Although
it may not be
the best criterion from the overall performance aspect \cite{Palomar2003}, the advantage of using Total-MSE is that
one can obtain
the optimal precoder structure or even the closed-form solution for the precoders in some cases (see \cite{RonghongMo2009, WeiGuan2008}). For the considered MIMO two-way relay system, we show that the closed-form relay precoder can be obtained under the Total-MSE criterion for given source precoders and decoders.

Before leaving this section, we provide some discussions on
the signalling overhead for obtaining
the CSI and the precoding information in the system.
First of all, we assume that the channel characteristics of each link change slowly enough so that
they can be perfectly estimated by using pilot symbols or training sequences.
If the channel reciprocity holds during the
MAC phase and BC phase  (e.g., they are in time-division duplex mode)
with ${\bf G}_1={\bf H}^T_1$ and ${\bf G}_2={\bf H}^T_2$, then the relay only needs to estimate the channel parameters during the MAC phase and the global CSI can be obtained. As a result, the joint precoding design can be conducted at the relay node and then the relay node broadcasts ${\bf A}_i$ to $S_i$, $i=1,2$. To cancel self-interference and demodulate the received signals, the source nodes should estimate the corresponding channel parameters. For example, $S_1$ needs to estimate ${\bf G}_1 {\bf A}_r {\bf H}_1$ to subtract the self-interference ${\bf s}_1$
and estimate ${\bf G}_1 {\bf A}_r {\bf H}_2$ to demodulate ${\bf s}_2$. If, on the other hand, the channel reciprocity does not hold during the MAC phase and BC phase (e.g., they are in frequency-division duplex mode), more
feedback channels and signalling
overheads are required. The relay can only estimate ${\bf H}_1$ and ${\bf H}_2$ during the MAC phase. To obtain the global CSI, the relay node needs $S_1$ and $S_2$ to feedback ${\bf G}_1$ and  ${\bf G}_2$, respectively.

\section{Iterative Precoding Design}
In this section, we first formulate the joint optimization of the source and relay precoding for Total-MSE minimization in the considered MIMO two-way relay systems. This problem is shown to be non-linear and non-convex and the optimal solution is not easily tractable. To approach the global optimal solution, we propose an iterative algorithm based on alternating optimization that updates one precoder at a time while fixing the others.

According to the received signal in \eqref{eqn:source1receivedsignalnewnew} and assuming linear receiver, the MSE at $S_i$ can be written as
\begin{equation} \label{eqn:MSE1}
    J_i={\cal E} \left \{ ||{\bf W}_i {\bf y}_i - {\bf s}_{\bar i} ||^2_2 \right \},~i=1,2
\end{equation}
where ${\bf W}_i \in {\mathbb C}^{N\times N}$ is the linear decoding matrix at the destination $S_i$. Substituting \eqref{eqn:source1receivedsignalnewnew} into \eqref{eqn:MSE1}, it further yields
\begin{equation}\label{eqn:MSE21}
\begin{split}
   J_i =&{\cal E} \left \{ ||{\bf W}_i \left({\bf F}_i  {\bf s}_{\bar i}
     + {\bf G}_i {\bf A}_r {\bf n}_r + {\bf n}_i \right)- {\bf s}_{\bar i} ||^2_2 \right \}\\
     =&{\rm Tr}\left\{ {\bf W}_i {\bf F}_i  {\bf F}^H_i  {\bf W}^H_i
     - {\bf W}_i {\bf F}_i
      -  {\bf F}^H_i  {\bf W}^H_i
     + \sigma^2_r {\bf W}_i {\bf G}_i {\bf A}_r  {\bf A}^H_r {\bf G}^H_i {\bf W}^H_i \right.\\
     &\left. + \sigma^2_i {\bf W}_i {\bf W}^H_i +{\bf I}_N \right\},~i=1,2
\end{split}
\end{equation}
where we have used the fact that ${\bf s}_i$, ${\bf n}_i$ and ${\bf n}_r$ are mutually independent.
The problem is to find the optimal precoding/decoding matrices $\left\{{\bf A}_r, {\bf A}_i, {\bf W}_i, i=1,2\right\}$ such that the Total-MSE of the two users can be minimized. This is formulated as
\begin{eqnarray}\label{OptimizationProb}
       && \min_{{\bf A}_r,{\bf A}_i, {\bf W}_i, i=1,2} J_1+J_2\\
      && s.t. ~~~\eqref{eqn:Powersource1}~~\eqref{eqn:RelayPowerCons} \nonumber
\end{eqnarray}
Before solving \eqref{OptimizationProb}, we present the following theorem.
Based on this theorem, we only consider the case $M\geq N$ throughout this paper.

\textbf{Theorem 1}:
When $M\geq N$, the Total-MSE $J_1+J_2$ can be made arbitrarily small by increasing the power at both the source nodes and the relay node in the considered $(N, M, N)$ two-way relay system. Otherwise if $M < N$, $J_1+J_2$ is always lower bounded by $2(N-M)$.
\begin{proof}
We first provide an alternative expression for the MSE of each source, $J_i$.
Since the constraints do not involve the decoding matrix ${\bf W}_i$ in the problem formulation \eqref{OptimizationProb}, a necessary condition for the optimal solution is $\frac{\partial J_i}{\partial {\bf W}^*_i}={\bf 0}$. By using the matrix differentiation rules in\cite{Hjorungnes2007}, the optimal solution of ${\bf W}_i$, denoted as ${\bf W}_i^{opt}$, can be expressed in closed-form as
\begin{equation}\label{OptimalW1}
   {\bf W}^{opt}_i={\bf F}^H_i {\bf R}^{-1}_{w_i},~i=1,2
\end{equation}
where
\begin{equation}\label{Rw1}
   {\bf R}_{w_i}={\bf F}_i {\bf F}^H_i + \sigma^2_r {\bf G}_i {\bf A}_r  {\bf A}^H_r {\bf G}^H_i+ \sigma^2_i {\bf I}_N.
\end{equation}
By substituting ${\bf W}_i^{opt}$ in \eqref{OptimalW1} into \eqref{eqn:MSE21}, the MSE at $S_i$, $J_i$, transforms into
\begin{equation} \label{NewJ1}
       {\hat J}_i = {\rm Tr}\left \{ \left[{\bf I}_N +
    {\bf F}^H_i \left (\sigma^2_i {\bf I}_N+ {\sigma}^2_r {\bf G}_i {\bf A}_r {\bf A}_r^H {\bf G}^H_i \right )^{-1}
   {\bf F}_i \right ]^{-1} \right\},~i=1,2.
\end{equation}
Therefore, the minimum Total-MSE $J_1+J_2$ of the original problem \eqref{OptimizationProb} will be the same as the minimum of
${\hat J}_1+{\hat J}_2$ subject to  the same power constraints.
For brevity of illustration, we take ${\hat J}_1$ as an example. Define ${\bf Q}=\left (\sigma^2_1 {\bf I}_N+ {\sigma}^2_r {\bf G}_1 {\bf A}_r {\bf A}_r^H {\bf G}^H_1 \right )^{-1}$ for simplicity of notation.
Note that the rank of
$\bf Q$ is equal to $N$. When $M\geq N$,
it is always possible to find precoders $\{{\bf A}_r,{\bf A}_2 \}$ to make the rank of
the term ${\bf F}^H_1 {\bf Q} {\bf F}_1$ equal to $N$.
Let ${a_n}$, $n=1,2,\cdots,N$, denote
the positive eigenvalues of ${\bf F}^H_1 {\bf Q} {\bf F}_1$, then ${\hat J}_1$ can be rewritten as
\begin{equation} \label{SVDhatJ1}
    {\hat J}_1 = {\rm Tr}\left \{ \left[{\bf I}_N +
    {\rm Diag}\left([a_1,a_2,\ldots,a_N ] \right )\right]^{-1} \right\}
    =\sum^N_{n=1}\frac{1}{1+a_n}.
\end{equation}
Next, we prove that
by increasing the power at both $S_2$ and $R$,
we can always increase $a_i$ and hence decrease the MSE ${\hat J}_1$.
Let us define
\begin{equation}\label{Theo1add1}\nonumber
\begin{split}
 {\bf E}&={\bf I}_N + {\bf F}^H_1  {\bf Q} {\bf F}_1 \\
&= {\bf I}_N + \theta_2 \theta_r \bar{{\bf A}}^H_2 {\bf H}^H_2 \bar{{\bf A}}^H_r {\bf G}^H_1\left (\sigma^2_1 {\bf I}_N+ \theta_r {\sigma}^2_r {\bf G}_1 \bar{{\bf A}}_r \bar{{\bf A}}_r^H {\bf G}^H_1 \right )^{-1}
{\bf G}_1 \bar{{\bf A}}_r {\bf H}_2 \bar{{\bf A}}_2,
\end{split}
\end{equation}
where we have replaced ${\bf F}_1$ by ${\bf G}_1 {\bf A}_r {\bf H}_2 {\bf A}_2$ as defined in \eqref{eqn:source1receivedsignalnewnew} when obtaining the second equation and set ${\bf A}_2=\sqrt{\theta_2}\bar{{\bf A}}_2$ and ${\bf A}_r=\sqrt{\theta_r}\bar{{\bf A}}_r$ with
$\theta_2$ and $\theta_r$ being power scalar parameters for ${\bf A}_2$ and ${\bf A}_r$, respectively.
Then, we can rewrite the MSE in \eqref{NewJ1} as ${\hat J}_1={\rm Tr} \{{\bf E} ^{-1} \}$.
It is easy to verify that enlarging $\theta_2$ can always increase the eigenvalues $a_i$ to decrease ${\hat J}_1$. However, due to the power constraint at the relay, we also need to check how $\theta_r$ affects ${\hat J}_1$. By defining $\beta=1/\theta_r$, we rewrite ${\bf E}$ as
\begin{equation}\label{Theo1add2}\nonumber
 {\bf E}={\bf I}_N + {\bf A}^H_2 {\bf H}^H_2 \bar{{\bf A}}^H_r {\bf G}^H_1\left (\beta \sigma^2_1 {\bf I}_N+ {\sigma}^2_r {\bf G}_1 \bar{{\bf A}}_r \bar{{\bf A}}_r^H {\bf G}^H_1 \right )^{-1}
{\bf G}_1 \bar{{\bf A}}_r {\bf H}_2 {\bf A}_2.
\end{equation}
Then, we have
\begin{equation}\nonumber
\begin{split}
      \frac{d{\rm Tr}({\bf E}^{-1})}{d\beta}&=
      {\rm Tr} \{-{\bf E}^{-1} d  [ {\bf A}^H_2 {\bf H}^H_2 \bar{{\bf A}}^H_r {\bf G}^H_1  (\underbrace{\sigma^2_r {\bf G}_1 \bar{{\bf A}}_r \bar{{\bf A}}_r^H {\bf G}^H_1 +\beta \sigma^2_1 {\bf I}_N }_{{\bf P}}  )^{-1} {\bf G}_1 \bar{{\bf A}}_r {\bf H}_2 {\bf A}_2 ]{\bf E}^{-1}  \}\\
      &={\rm Tr}\left\{\sigma^2_1  {\bf E}^{-1} {\bf A}^H_2 {\bf H}^H_2 \bar{{\bf A}}^H_r {\bf G}^H_1 {\bf P}^{-2}
      {\bf G}_1 \bar{{\bf A}}_r {\bf H}_2 {\bf A}_2 {\bf E}^{-1}\right\}> 0,
\end{split}
\end{equation}
where we have used the fact that both ${\bf E}$ and ${\bf P}$ are positive definite. Therefore,
we conclude that ${\hat J}_1$ is a monotonically decreasing function with respect to $\theta_r$.
It suggests that enlarging $\theta_r$ can also increase $a_i$ and decrease ${\hat J}_1$.

Secondly, we show that if $M \geq N$, the MSE $J_i$ can be made arbitrarily small by increasing the power at both source and relay nodes. To this end, we simply assume that when increasing the power at $S_2$ (i.e., increasing the scalar $\theta_2$), the relay just increases its power to keep $\theta_r$ unchanged. Thus, similar to \eqref{SVDhatJ1}, we have
\begin{equation} \label{add1}
    {\hat J}_1
    =\sum^N_{n=1}\frac{1}{1+\theta_2 \bar{a}_n},
\end{equation}
where $\bar{a}_n, n=1,2,\cdots,N$, are the eigenvalues of $\bar{{\bf A}}^H_2 {\bf H}^H_2 {\bf A}^H_r {\bf G}^H_1 {\bf Q} {\bf G}_1 {\bf A}_r {\bf H}_2 \bar{{\bf A}}_2 $.
For an arbitrarily small $\epsilon_1$, by
defining $\bar{a}_{min}=\min\{\bar{a}_1,\bar{a}_2,\cdots,\bar{a}_N\}$,
we can always have
 \begin{equation} \label{add2}
\begin{split}
    {\hat J}_1
    &=\sum^N_{i=1}\frac{1}{1+\theta_2 \bar{a}_i}\\
    &\leq \frac{N}{1+\theta_2 \bar{a}_{min}} \leq \epsilon_1
\end{split}
\end{equation}
if $\theta_2\geq \frac{N/\epsilon_1-1}{\bar{a}_{min}}$.

On the other hand, if $M< N$, the maximum rank of the term
${\bf F}^H_1 {\bf Q} {\bf F}_1$
in ${\hat J}_1$ is $M$.
Assuming that the $M$ non-zero eigenvalues of ${\bf F}^H_1 {\bf Q} {\bf F}_1$ are denoted by $\{b_1,b_2,\cdots,b_M\}$,
the resultant ${\hat J}_1$ can be expressed as
\begin{equation} \label{newadd}\nonumber
    {\hat J}_1 = \sum^{M}_{n=1}\frac{1}{1+b_n}+(N-M).
\end{equation}
No matter how much power is provided at the source and relay nodes,
${\hat J}_1$ is always lower bounded by $N-M$.
The same
bound holds for ${\hat J}_2$. Theorem 1 is thus proven.
\end{proof}

We now take a closer look at the problem \eqref{OptimizationProb},
which can be proven to be
non-linear and non-convex and hence is difficult to solve. To make the problem tractable,
we propose an iterative algorithm which decouple the primal problem into three sub-problems and solve each of them in an alternating optimization approach.

First, given the precoding matrices at the source and relay nodes, i.e., ${\bf A}_1$, ${\bf A}_2$ and ${\bf A}_r$, we try to find the optimal decoder matrices ${\bf W}_1$ and ${\bf W}_2$.
Since the power constraints in \eqref{eqn:Powersource1} and \eqref{eqn:RelayPowerCons} are not related to ${\bf W}_1$ and ${\bf W}_2$, we simply get an unconstrained optimization problem
\begin{equation}\label{Optimization1}
     \min_{ {\bf W}_1, {\bf W}_2} J_{w_1}+J_{w_2}
\end{equation}
where $J_{w_i}={\rm Tr} \left\{ {\bf W}_i {\bf R}_{w_i} {\bf W}^H_i -{\bf W}_i {\bf F}_i
     -  {\bf F}^H_i  {\bf W}^H_i + {\bf I}_N \right\},~i=1,2$,
with ${\bf R}_{w_i}$ given in \eqref{Rw1}.
Since ${\bf R}_{w_i}$, $i=1,2$, is positive definite,
the objective function in \eqref{Optimization1} is convex with respect to ${\bf W}_i$.
Therefore,
applying the Karush-Kuhn-Tucker (KKT) conditions,
we obtain the optimal decoding matrices as described in \eqref{OptimalW1} by equating the gradient of objective function in \eqref{Optimization1} to zero.

Second, we consider the optimization of the relay precoding matrix ${\bf A}_r$ by assuming that ${\bf W}_i$, ${\bf A}_i$, $i=1,2$, are fixed. From \eqref{eqn:MSE21},
this sub-problem is equivalent to
\begin{eqnarray}\label{Optimization2}
    && \min_{ {\bf A}_r } ~J_{r_1}+J_{r_2} \\
    s.t.~&& {\rm Tr} \left\{ {\bf A}_r {\bf R}_x {\bf A}^H_r \right\} \leq \tau_r \IEEEyessubnumber \label{Optimiztion21}
\end{eqnarray}
where
$J_{r_i}$ is obtained by replacing ${\bf F}_i$ in \eqref{eqn:MSE21} with ${\bf G}_i {\bf A}_r {\bf H}_{\bar i}{\bf A}_{\bar i}$ as defined in \eqref{eqn:source1receivedsignalnewnew} and using the circular property of trace operator ${\rm Tr}\{{\bf A}{\bf B}\}={\rm Tr}\{{\bf B}{\bf A}\}$, given by
\begin{equation} \label{add3}
\begin{split}
J_{r_i}=& {\rm Tr} \left\{ {\bf G}^H_i {\bf W}^H_i {\bf W}_i {\bf G}_i {\bf A}_r {\bf R}_{x_{\bar i}} {\bf A}^H_r
    - {\bf H}_{\bar i} {\bf A}_{\bar i} {\bf W}_i {\bf G}_i {\bf A}_r \right. \\
     & \left. - {\bf G}^H_i {\bf W}^H_i  {\bf A}^H_{\bar i} {\bf H}^H_{\bar i} {\bf A}^H_r
    + \sigma^2_i {\bf W}_i {\bf W}^H_i +{\bf I}_N \right\},~i=1,2
\end{split}
\end{equation}
with
${\bf R}_{x_i}={\bf H}_i {\bf A}_i {\bf A}^H_i {\bf H}^H_i+\sigma^2_r {\bf I}_{M}$,
and \eqref{Optimiztion21} refers to the relay power constraint defined in \eqref{eqn:RelayPowerCons} with
\begin{equation} \nonumber
{\bf R}_x={\bf H}_1 {\bf A}_1 {\bf A}^H_1 {\bf H}^H_1+{\bf H}_2 {\bf A}_2 {\bf A}^H_2 {\bf H}^H_2+\sigma^2_r {\bf I}_{M}.
\end{equation}
Note the source power constraints \eqref{eqn:Powersource1} are irrelevant here since ${\bf A}_1$ and ${\bf A}_2$ are fixed.

\textbf{Lemma 1}:
The problem of relay precoding design given source precoders and decoders for Total-MSE minimization in the considered $(N, M, N)$ MIMO two-way relay system as formulated in \eqref{Optimization2} is convex.
\begin{proof}
Please refer to Appendix~\ref{prof_lemma1}.
\end{proof}

Due to the convexity of the problem \eqref{Optimization2}, we can readily design the optimal relay precoder by employing the KKT conditions. Specifically, the Lagrangian function of \eqref{Optimization2} is given as
\begin{equation}\label{Lagrangian}\nonumber
    {\cal L}=J_{r_1}+ J_{r_2}+ \lambda \left({\rm Tr} \left\{ {\bf A}_r {\bf R}_x {\bf A}^H_r \right\} - \tau_r \right),
\end{equation}
where $\lambda\geq0$ is the Lagrangian multiplier. Thus, the KKT conditions are
\begin{equation}\label{KKTconditions1}
    \frac{\partial {\cal L}}{\partial {\bf A}^{*}_r}=
    {\bf R}_{r_1} {\bf A}_r {\bf R}_{x_2}
    +  {\bf R}_{r_2} {\bf A}_r {\bf R}_{x_1} - {\bf R}_r
    +\lambda {\bf A}_r {\bf R}_x ={\bf 0},
\end{equation}
\begin{equation}\label{KKTconditions2}
   \lambda \left({\rm Tr} \left\{ {\bf A}_r {\bf R}_x {\bf A}^H_r \right\} - \tau_r \right)=0,~
\end{equation}
\begin{equation}\label{KKTconditions3}
    {\rm Tr} \left\{{\bf A}_r {\bf R}_x {\bf A}^H_r \right\} \leq \tau_r,
\end{equation}
where
${\bf R}_r={\bf G}^H_1 {\bf W}^H_1 {\bf A}^H_2 {\bf H}^H_2 + {\bf G}^H_2 {\bf W}^H_2 {\bf A}^H_1 {\bf H}^H_1$
and
${\bf R}_{r_i}={\bf G}^H_i {\bf W}^H_i {\bf W}_i {\bf G}_i$, $i=1,2$. To obtain \eqref{KKTconditions1}, the differentiation rule $\frac{\partial {\rm Tr}\{ {\bf Z}{\bf A}_0 {\bf Z}^H {\bf A}_1\}}{\partial {\bf Z}^{*}}={\bf A}_1 {\bf Z}{\bf A}_0$ in \cite{Hjorungnes2007} is applied.

Based on \eqref{KKTconditions1} we further obtain
\begin{equation}\label{Ar2}
    {\bf A}^{opt}_r=mat\left\{ \left[{\bf R}^T_{x_2}\otimes {\bf R}_{r_1} + {\bf R}^T_{x_1} \otimes {\bf R}_{r_2}
    + \lambda {\bf R}^T_x \otimes {\bf I}_{M}
    \right]^{-1} vec({\bf R}_r) \right\}.
\end{equation}
In the special case when $\lambda = 0$, we have
\begin{equation}\label{add4}
    {\bf A}^{opt}_r=mat\left\{ \left[{\bf R}^T_{x_2}\otimes {\bf R}_{R_1} + {\bf R}^T_{x_1} \otimes {\bf R}_{R_2}
    \right]^{-1} vec({\bf R}_r) \right\}.
\end{equation}
If ${\bf A}^{opt}_r$ in \eqref{add4} meets the condition \eqref{KKTconditions3}, then \eqref{add4} is the optimal relay precoder.
Otherwise, $\lambda$ in \eqref{Ar2} should be chosen to satisfy $ {\rm Tr} \left\{{\bf A}_r {\bf R}_x {\bf A}^H_r \right\} = \tau_r$.

\textbf{Lemma 2}: The function $g(\lambda)={\rm Tr} \left\{ {\bf A}_r {\bf R}_x {\bf A}^H_r \right\}$, with ${\bf A}_r$ given by \eqref{Ar2}, is monotonically decreasing with respect to $\lambda$ and the optimal $\lambda$ is upper-bounded by $\sqrt{\frac{{\bf R}_r {\bf R}^{-1}_x {\bf R}^H_r } {\tau_r}}$.

\begin{proof}
Please refer to Appendix~\ref{prof_lemma2}.
\end{proof}
With Lemma 2, the
optimal $\lambda$
meeting the condition $ {\rm Tr} \left\{{\bf A}_r {\bf R}_x {\bf A}^H_r \right\} = \tau_r$
can be readily obtained using bisection search.

The third sub-problem is to optimize the source precoder ${\bf A}_i$ for fixed ${\bf A}_r$ and ${\bf W}_i$, $i=1,2$.
This is formulated as:
\begin{eqnarray}\label{Optimiztion3}
       && \min_{{\bf A}_1,{\bf A}_2}  J_{s_1}+J_{s_2} \\ \nonumber
       s.t.
       && {\rm Tr}\left\{ {\bf A}_i {\bf A}^H_i \right\} \leq \tau_i,~i=1,2 \\ \nonumber
       && {\rm Tr}\left\{ {\bf R}_{p_1}{\bf A}_1 {\bf A}^H_1
       + {\bf R}_{p_2} {\bf A}_2 {\bf A}^H_2 \right\}
       \leq \tau^{'}_r \IEEEyessubnumber \label{Optimiztion31}
\end{eqnarray}
where $\tau^{'}_r =\tau_r- \sigma^2_r {\rm Tr}\left\{ {\bf A}_r {\bf A}^H_r \right\}$, ${\bf R}_{p_i}= {\bf H}^H_i {\bf A}^H_r {\bf A}_r {\bf H}_i$, $i=1,2$ and
\begin{equation} \label{add5}
J_{s_i}={\rm Tr}\left\{ {\bf R}_{s_{i1}} {\bf A}_{\bar i} {\bf A}^H_{\bar i} -2 {\Re} \left( {\bf R}_{s_{i2}} {\bf A}_{\bar i} \right) + {\bf R}_{s_{i3}} \right\}, i=1,2
\end{equation}
with
\begin{equation} \nonumber
{\bf R}_{s_{i1}}= {\bf H}^H_{\bar i} {\bf A}^H_r {\bf G}^H_i {\bf W}^H_i {\bf W}_i {\bf G}_i {\bf A}_r {\bf H}_{\bar i},
\end{equation}
\begin{equation} \nonumber
{\bf R}_{s_{i2}}= {\bf W}_i {\bf G}_i {\bf A}_r {\bf H}_{\bar i},
\end{equation}
\begin{equation} \nonumber
{\bf R}_{s_{i3}}=\sigma^2_r {\bf W}_i {\bf G}_i {\bf A}_r  {\bf A}^H_r {\bf G}^H_i {\bf W}^H_i+ \sigma^2_i {\bf W}_i {\bf W}^H_i +{\bf I}_N.
\end{equation}
To obtain \eqref{add5}, the circular property of trace operator is again applied for \eqref{eqn:MSE21}.

It is noted that the change of source precoders can affect the power constraint at the relay. Hence, the relay power constraint should be included as \eqref{Optimiztion31} in \eqref{Optimiztion3}. By applying the conclusion derived in
Lemma A (given in Appendix~\ref{prof_lemma1}),
we can also prove that the optimization problem \eqref{Optimiztion3} is convex.

\textbf{Lemma 3}:
The optimization problem in the form of \eqref{Optimiztion3} can be transformed into a convex quadratically constrained quadratic program (QCQP) problem.
\begin{proof}
Please refer to Appendix~\ref{prof_Optimiztion3}.
\end{proof}
A QCQP problem can be efficiently solved by applying the available software package \cite{CVX}.

In summary, we outline the iterative precoding design algorithm as follows:

\vspace{-0.3cm}
\hrulefill
\par
{\footnotesize
\textbf{Algorithm 1} (Iterative precoding)
\begin{itemize}
\item \textbf{Initialize} ${\bf A}_1$, ${\bf A}_2$ and ${\bf A}_r$\footnote{Here, ${\bf A}_i$ and ${\bf A}_r$ can be randomly generated complex matrices
or set as identity matrices, as long as they satisfy the given power constraints.}
\item \textbf{Repeat}
\begin{itemize}
\item Update the decoder matrices ${\bf W}_1$ and ${\bf W}_2$ using \eqref{OptimalW1} for fixed ${\bf A}_1$, ${\bf A}_2$ and ${\bf A}_r$;
\item Update the relay precoder matrices ${\bf A}_r$ using \eqref{Ar2} or \eqref{add4} for fixed ${\bf A}_1$, ${\bf A}_2$, ${\bf W}_1$ and ${\bf W}_2$;
\item For fixed ${\bf A}_r$, ${\bf W}_1$ and ${\bf W}_2$, solve the convex QCQP problem to get the optimal ${\bf A}_1$ and ${\bf A}_2$ as in Appendix~\ref{prof_Optimiztion3};
\end{itemize}
\item \textbf{Until} termination criterion is satisfied.
\end{itemize}}
\vspace{-0.5cm}
\hrulefill

\textbf{Theorem 2}: The proposed iterative precoding design algorithm, Algorithm 1, is convergent and the limit point of the iteration is a stationary point of
\eqref{OptimizationProb}.

\begin{proof}
Since in the proposed algorithm, the solution for each subproblem is optimal, the Total-MSE is decreased with each iteration. Meanwhile, the Total-MSE is lower bounded (at least by zero). Hence, the proposed algorithm is convergent.
It further means that there must exist a
limit point, denoted as $\left\{ {\bf \bar{W}}_i, {\bf \bar{A}}_i, i=1,2, {\bf \bar{A}}_r\right\}$, after the convergence. At the limit point, the solutions will not change if we continue the iteration. Otherwise, the Total-MSE can be further decreased and it contradicts the assumption of convergence. Since ${\bf \bar{W}}_i$, ${\bf \bar{A}}_i$ ($i=1,2$) and ${\bf \bar{A}}_r$ are local minimizers for each subproblem, we have
\begin{equation}\label{StationaryPoint1}\nonumber
    {\rm Tr}\{{\triangledown}_{{\bf W}_i} {J_w}({\bf \bar{W}}_i; {\bf \bar{A}}_i,  {\bf \bar{A}}_r, i=1,2)^T ( {\bf W}_i-{\bf \bar{W}}_i) \}\geq 0,
\end{equation}
\begin{equation}\label{StationaryPoint2}\nonumber
    {\rm Tr}\{{\triangledown}_{{\bf A}_r} {J_r}({\bf \bar{A}}_r; {\bf \bar{A}}_i,{\bf \bar{W}}_i,i=1,2 )^T ( {\bf A}_r-{\bf \bar{A}}_r) \}\geq 0,
\end{equation}
\begin{equation}\label{StationaryPoint3}\nonumber
    {\rm Tr}\{{\triangledown}_{{\bf A}_i} {J_s}({\bf \bar{A}}_i; {\bf \bar{W}}_i, {\bf \bar{A}}_r, i=1,2 )^T ( {\bf A}_i-{\bf \bar{A}}_i) \}\geq 0,
\end{equation}
where $J_w=J_{w_1}+J_{w_2}$, $J_r=J_{r_1}+J_{r_2}$ and $J_s=J_{s_1}+J_{s_2}$. Summing up all the above equations, we get
\begin{equation}\label{StationaryPoint4}
    {\rm Tr}\{{\triangledown}_{{\bf X}} {J}({\bf \bar{X}} )^T  ( {\bf X}-{\bf \bar{X}} )  \}\geq 0,
\end{equation}
where $J=J_1+J_2$ and ${\bf X}=\left[{\bf W}_1, {\bf W}_2, {\bf A}_1, {\bf A}_2, {\bf A}_r\right]$. Result \eqref{StationaryPoint4} implies the stationarity of ${\bf \bar{X}}$ of \eqref{OptimizationProb} by definition.
\end{proof}

\textit{Remark 1}: In this work, the precoders are designed to minimize the Total-MSE of all the data streams of two users. This may lead to
unbalanced MSE distribution among the data streams. In general, the overall error performance is dominated
by the data stream with the highest MSE \cite{Palomar2003}. Therefore, an alternative objective is to minimize
the maximum per-stream MSE among all the data streams in order to improve the overall performance.
Nevertheless, in \cite{Palomar2003}, it has been proven that the min-max MSE problem can be solved through the Total-MSE minimization. Specifically, the solutions to the min-max problem can be obtained by multiplying the source precoder ${\bf A}_i$ of the Total-MSE problem with a rotation matrix to make MSE matrix with equal diagonal entries.

\section{Low-Complexity Precoding Design Based on Channel Parallelization}
The iterative precoding design algorithm presented in Section III obtains good performance as verified in Section VI, but also has high computational complexity. In this section, we propose a new precoding design that offers a good balance between performance and complexity.

It has been proven in \cite{ChunguoLi2009, RonghongMo2009,RonghongMo2009a,WeiGuan2008,Tseng2009,ZhengFang2006} that the optimal precoding structure in one-way relaying is to first parallelize the channels between the source and the relay,
as well as between the relay and the destination  using singular value decomposition (SVD) and then match the eigen-channels in the two hops.
Taking the transmission of single data stream in a one-way relay system for example as considered in \cite{Khoshnevis2008a} and \cite{Havary-Nassab2009},
the idea of channel matching is as follows.
The source should use the dominant right singular vector of the channel in the first hop as beamformer to transmit its signal. After receiving the signal from the source, the relay should first multiply it with the dominant left singular vector of the same channel and then transmit it through the dominant right singular vector of the channel in the second hop.

Motivated by the findings in \cite{ChunguoLi2009, RonghongMo2009,RonghongMo2009a,WeiGuan2008,Tseng2009,ZhengFang2006,Khoshnevis2008a, Havary-Nassab2009},
we aim to design ${\bf A}_1$, ${\bf A}_2$ and ${\bf A}_r$ so as to simultaneously parallelize the bidirectional links in the MIMO two-way relay system.
In the following, we introduce a heuristic channel parallelization method for bidirectional communications by using two joint channel decomposition methods, namely, generalized singular value decomposition (GSVD) for the MAC phase and SVD for the BC phase.
Using this method we then reduce
the precoder design to a simple power allocation problem.

\subsection{ Channel Parallelization}
The major task
of simultaneously parallelizing the bidirectional links
is to
jointly decompose the forward channel matrix pair $\{{\bf H}_1, {\bf H}_2\}$ in the MAC phase
and the backward channel matrix pair $\{{\bf G}_1, {\bf G}_2\}$ in the BC phase.
To do so, we first apply the GSVD technique for the MAC channels.
The GSVD is elaborated in the lemma below.

\textbf{Lemma 4}\cite{xiandazhang2004}: Assuming ${\bf A}\in {\mathbb C}^{m\times n}$ and ${\bf B}\in {\mathbb C}^{m\times n}$, $m \leq n \leq 2m$ are two full-rank matrices that satisfy ${\rm Rank}\left[ \begin{array}{c}
                      {\bf A} \\
                      {\bf B}
                    \end{array}\right]=n$, there exist two ${m\times m}$ unitary matrices ${\bf U}_A$, ${\bf U}_B$ and an ${n\times n}$ non-singular matrix ${\bf V}$ which make
\begin{equation} \label{add7} \nonumber
\begin{split}
 {\bf A}  = {\bf U}_A \bm{\Sigma}_A {\bf V},~
 {\bf B}={\bf U}_B \bm{\Sigma}_B  {\bf V} ,
\end{split}
\end{equation}
where $\bm{\Sigma}_A = [{\bf 0}_{m \times (n-m)}, \bm{\Lambda}_A]$, $\bm{\Sigma}_B = [\bm{\Lambda}_B, {\bf 0}_{m \times (n-m)}]$ and they satisfy $\bm{\Sigma}^T_A \bm{\Sigma}_A + \bm{\Sigma}^T_B \bm{\Sigma}_B= {\bf I}_n$. Here $\bm{\Lambda}_A$ and $\bm{\Lambda}_B$ are two $m \times m$ non-negative diagonal matrices.

By applying Lemma 4 onto the channel pair $\{{\bf H}^H_1,{\bf H}^H_2\}$,
${\bf H}_1$ and ${\bf H}_2$ can be expressed as\footnote{To apply Lemma 4, we here assume that $M\leq 2N$.}
\begin{equation} \label{eqn:GSVDHnew}
    {\bf H}_1={\bf V}_h {\bf \Sigma}_{h_1} { \bf U}^H_{h_1},~~~
    {\bf H}_2={\bf V}_h {\bf \Sigma}_{h_2} {\bf U}^H_{h_2},
\end{equation}
where ${\bf V}_h$ is a non-singular $M \times M$ complex matrix, ${ \bf U}_{h_1}$ and ${ \bf U}_{h_2}$ are two $N\times N$ unitary matrices, ${\bf \Sigma}_{h_1}=\left[{\bf 0}^T_{(M-N)\times N},{\bf \Lambda}^T_{h_1} \right]^T$ and ${\bf \Sigma}_{h_2}=\left[{\bf \Lambda}^T_{h_2}, {\bf 0}^T_{(M-N)\times N} \right]^T$ where ${\bf \Lambda}_{h_1}$ and
${\bf \Lambda}_{h_2}$ are two $N\times N$ non-negative diagonal matrices.
If the relay precoder ${\bf A}_r$ contains ${{\bf V}_h}^{-1}$ at the right side and ${\bf A}_i$ has ${\bf U}_{h_i}$ at the left side, we can parallelize the two forward channels in the MAC phase.

For the BC phase, since the superimposed signal should be simultaneously transmitted to two destinations, we construct one virtual point-to-point MIMO channel as ${\bf G}=\left[ {\bf G}^T_1, {\bf G}^T_2 \right]^T$. By imposing SVD technique on $\bf G$, we have
\begin{equation} \label{eqn:GSVDG}
    {\bf G}={\bf V}_g {\bf \Sigma }_g {\bf U}^H_g,
\end{equation}
where ${\bf V}_g$ and ${\bf U}_g$ are $2N\times2N$ and $M\times M$ unitary matrices, respectively.
${\bf\Sigma}_g=\left[ {\bf\Lambda}^T_g,  {\bf 0}^T_{(2N-M)\times M} \right]^T$
where ${\bf\Lambda}_g$ is an $M\times M$ non-negative diagonal matrix.
If ${\bf A}_r$ contains ${\bf U}_g$ at its left side, the virtual point-to-point MIMO channel $\bf G$ is parallelized in the BC phase.
Accordingly, we can rewrite ${\bf G}_1$ and ${\bf G}_2$ as
\begin{equation} \label{eqn:GSVDG1} \nonumber
    {\bf G}_1={\bf V}_{g_1} {\bf\Sigma}_g {\bf U}^H_g, ~~~
    {\bf G}_2={\bf V}_{g_2} {\bf\Sigma}_g {\bf U}^H_g,
\end{equation}
where ${\bf V}_{g_1}={\bf V}_g(1:N,1:2N)$ and ${\bf V}_{g_2}={\bf V}_g(N+1:2N,1:2N)$. Note that ${\bf V}_{g_1}$ and ${\bf V}_{g_2}$ no longer have the unitary property.

We now readily propose the following structure for the three precoders:
\begin{equation} \label{eqn:A1}
    {\bf A}_1={\bf U}_{h_1}  {\bf \Lambda}_{A_1}{\bf V}_{A_1}, ~~{\bf A}_2={\bf U}_{h_2} {\bf \Lambda}_{A_2}{\bf V}_{A_2}, ~~
    {\bf A}_r={\bf U}_{g} {\bf \Lambda}_{A_r} {\bf V}^{-1}_{h},
\end{equation}
where ${\bf V}_{A_1}$ and ${\bf V}_{A_2}$ are arbitrary unitary matrices, ${\bf \Lambda}_{A_1}$, ${\bf \Lambda}_{A_2}$ and ${\bf \Lambda}_{A_r}$ are $N\times N$, $N\times N$ and $M\times M$ real diagonal matrices, respectively, to be optimized in the next subsection.

The received signal in \eqref{eqn:source1receivedsignalnewnew} can therefore be rewritten as
\begin{equation} \label{Paray1}
    {\bf y}_i ={\bf V}_{g_i} {\bf \Sigma}_{g} {\bf \Lambda}_{A_r} {\bf \Sigma}_{h_{\bar i}} {\bf \Lambda}_{A_{\bar i}} {\tilde {\bf s}}_{\bar i}
     + {\bf V}_{g_i} {\bf \Sigma}_{g} {\bf \Lambda}_{A_r} {\tilde{\bf n}}_r + {\bf n}_i,~i=1,2
\end{equation}
where ${\tilde {\bf s}}_i={\bf V}_{A_i} {\bf s}_i$ and ${\tilde{\bf n}}_r={\bf V}^{-1}_{h} {\bf n}_r$. Note that ${\bf V}_{A_i}$ being unitary, it does not affect the statistical property of ${\bf s}_i$ nor the designed precoders.
Given $M > N$, since ${\bf \Sigma}_{h_1}=\left[{\bf 0}^T_{(M-N)\times N},{\bf \Lambda}^T_{h_1} \right]^T$ and ${\bf \Sigma}_{h_2}=\left[{\bf \Lambda}^T_{h_2}, {\bf 0}^T_{(M-N)\times N} \right]^T$
as given by the GSVD, the effective channel gains for the $N$ data streams of two sources can not be matched simultaneously. In other words, the gain of a certain data stream for $S_1$ may be very strong while the gain of the corresponding data stream for $S_2$ can be very weak.
To avoid such unbalance, if not specified otherwise, we only consider the case with $M=N$ where all the channel gains can be utilized for
transmission of both users in the following of this section.
Then, \eqref{Paray1} turns to
\begin{equation} \label{Paray1New}\nonumber
    {\bf y}_i ={\bf{ \tilde V}}_{g_i} {\bf \Lambda}_{g} {\bf \Lambda}_{A_r} {\bf \Lambda}_{h_{\bar i}} {\bf \Lambda}_{A_{\bar i}} {\tilde {\bf s}}_{\bar i}
     + {\bf{ \tilde V}}_{g_i} {\bf \Lambda}_{g} {\bf \Lambda}_{A_r} {\tilde{\bf n}}_r + {\bf n}_i,~i=1,2
\end{equation}
where ${\bf{ \tilde V}}_{g_1}={\bf V}_g(1:N,1:N)$ and ${\bf{ \tilde V}}_{g_2}={\bf V}_g(N+1:2N,1:N)$, and ${\bf \Lambda}_k$, for $k \in \{A_1, A_2, A_r,g, h_1,h_2\}$, is an $
N\times N$ non-negative diagonal matrix.
\subsection{Joint Power Allocation}
Based on the precoder structures proposed in \eqref{eqn:A1}, in this subsection we discuss the joint optimization of ${\bf \Lambda}_{A_1}$, ${\bf \Lambda}_{A_2}$ and ${\bf \Lambda}_ {A_r}$ to minimize the Total-MSE of the two users.
By substituting \eqref{eqn:A1} into \eqref{NewJ1}, we rewrite ${\hat J}_i$ as
\begin{equation} \label{eqn:MSEnew1}
   {\hat J}_i =
   {\rm Tr}\left\{ \left [ {\bf I}_{N} +
   \left(   {\bf \Lambda}_{A_{\bar i}} {\bf \Lambda}_{h_{\bar i}} {\bf \Lambda}_{A_r}{\bf \Lambda}_{g} \right)
   \left (\sigma^2_i {\bf B}_{g_i}+
   {\sigma}^2_r {\bf \Lambda}_{g} {\bf \Lambda}_{A_r}{\bf B}_{h}{\bf \Lambda}_{A_r} {\bf \Lambda}_{g}  \right )^{-1}
   \left( {\bf \Lambda}_{g} {\bf \Lambda}_{A_r} {\bf \Lambda}_{h_{\bar i}} {\bf \Lambda}_{A_{\bar i}} \right)\right ]^{-1}\right\},
\end{equation}
where ${\bf B}_{g_i}= \left({\bf{ \tilde V}}^H_{g_i} {\bf{ \tilde V}}_{g_i}\right)^{-1}$ and ${\bf B}_h= \left({\bf V}^H_h {\bf V}_h\right)^{-1}$.
It is found that, although ${\hat J}_i$, $i=1,2$ has been simplified, the MSE covariance matrices are still non-diagonal. Solving the optimization problem directly becomes difficult. However, we can resort to a tractable upper bound on the MSE to simplify the problem.

\textbf{Lemma 5}:
An upper bound of
${\hat J}_i$ defined in \eqref{eqn:MSEnew1} is given by
\begin{equation} \label{eqn:upperJ1}
   {\hat J}_i \leq {\rm Tr} \left\{ \left [ {\bf I}_{N} +
   \left(   {\bf \Lambda}_{A_{\bar i}} {\bf \Lambda}_{h_{\bar i}} {\bf \Lambda}_{A_r} {\bf \Lambda}_{g} \right)
   \left (\sigma^2_i {\bf \Lambda}_{Bg_i}+
   {\sigma}^2_r {\bf \Lambda}_{g} {\bf \Lambda}_{A_r} {\bf \Lambda}_{Bh} {\bf \Lambda}_{A_r} {\bf \Lambda}_{g}   \right )^{-1}
   \left( {\bf \Lambda}_{g} {\bf \Lambda}_{A_r} {\bf \Lambda}_{h_{\bar i}} {\bf \Lambda}_{A_{\bar i}} \right)  \right ]^{-1} \right\},
\end{equation}
where ${\bf \Lambda}_{Bg_i}$ and ${\bf \Lambda}_{Bh}$ are two diagonal matrices that contain the diagonal entries of
${\bf B}_{g_i}$ and ${\bf B}_h$, respectively.
\begin{proof}
Please refer to Appendix~\ref{prof_Lemma4}.
\end{proof}

The MSE
upper bound matrix in \eqref{eqn:upperJ1}
has a diagonal structure.
Therefore, we can minimize the upper bound to design the precoders. By further assuming
${\bf P}_k={\bf \Lambda}^2_k$ for $k \in \{A_1, A_2, A_r,g, h_1,h_2\}$,
the upper bound in Lemma 5 denoted as $J^u_i$ can be reformulated as
\begin{equation}\label{eqn:J1new}
   J^u_i=\sum^N_{n=1} \left( 1+ \frac{p^n_{g} p^n_{A_r} p^n_{h_{\bar i}} p^n_{A_{\bar i}}} {\sigma_i^2 \lambda^n_{Bg_i} +\sigma^2_r \lambda^n_{Bh} p^n_{g} p^n_{A_r}} \right)^{-1},~i=1,2
\end{equation}
where $p^n_k$'s are the diagonal entries of ${\bf P}_k$
and $\lambda^n_k$'s with $k\in \{Bh, Bg_1, Bg_2\}$ are the diagonal entries of ${\bf \Lambda}_k$.
It is interesting to find that $J^u_i$ is the Total-MSE of each sub-parallelized channel after zero forcing ${\bf y}_i$ by ${\bf V}^{-1}_{g_i}$.

Finally, the precoder design can  be simplified to the optimization problem as follows:
\begin{eqnarray} \label{eqn:upperprobelem}
    &&\min_{ p^n_{A_1}, p^n_{A_2}, p^n_{A_r}, \forall n }
    J^u_1 +J^u_2\\
    {s.t.}&&  \sum^N_{n=1} {p^n_{A_1}} \leq \tau_1, \nonumber
    ~\sum^N_{n=1} {p^n_{A_2} } \leq \tau_2,~p^n_{A_1} \geq 0,~p^n_{A_2} \geq 0,~p^n_{A_r} \geq 0 \\
    && \sum^N_{n=1} p^n_{A_r} \left ( p_{h_1}^n p^n_{A_1} + p_{h_2}^n p^n_{A_2} + \sigma^2_r {\lambda^n_{Bh}} \right ) \leq \tau_r \nonumber
\end{eqnarray}
Compared with the original objective function in \eqref{eqn:MSEnew1}, the expression in \eqref{eqn:upperprobelem} exhibits a simpler form and is more analytically tractable.
Nevertheless, the problem \eqref{eqn:upperprobelem} is still a non-convex optimization problem. In the following, we apply the iterative approach to convert the problem \eqref{eqn:upperprobelem} into two convex sub-problems.
\subsubsection{Sub-problem~1}
For given $p^n_{A_1}$ and $p^n_{A_2}$, $\forall n$, we
formulate the following problem
as follows to get the optimal ${\bf P}_{A_r}$
\begin{eqnarray} \label{eqn:subprobelem1}
   && \min_{p^n_{Ar}, \forall n }  J^u_1 + J^u_2 \\ \nonumber
   {s.t.} &&~ \sum^N_{n=1} p^n_{A_r} \left ( p^n_{h_1} p^n_{A_1} + p^n_{h_2} p^n_{A_2} + \sigma^2_r {\lambda^n_{Bh}} \right ) \leq \tau_r,~
    {\rm p}^n_{A_r} \geq 0,~\forall n \nonumber
\end{eqnarray}
By verifying
\begin{equation} \label{eqn:partialJ1Ar} \nonumber
    \frac{\partial^2 J^u_i}{\partial {p^n_{A_r}}^2}=
    \frac{2\sigma^2_i \lambda^n_{Bg_i} p^n_{h_{\bar i}} p^n_g p^n_{A_{\bar i}} \left(\sigma^2_r \lambda^n_{Bh} p^n_g + p^n_g p^n_{h_{\bar i}} p^n_{A_{\bar i}}\right)}
    {\left[\sigma^2_i \lambda^n_{Bg_i} + p^n_{A_r} \left(\sigma^2_r \lambda^n_{Bh} p^n_g + p^n_g p^n_{h_{\bar i}} p^n_{A_{\bar i}}\right)\right]^3} >0,~i=1,2
\end{equation}
we conclude that this sub-problem is convex.
Based on the KKT conditions (details presented in Appendix~\ref{prof_PowerAr}), we derive the water-filling solution
\begin{equation} \label{eqn:KKT8}
    p^n_{A_r} =\max \left[0, \textit{{\rm Root}}(f)\right], ~~\forall n
\end{equation}
where $\textit {\rm Root}(f)$ denotes the maximum real root of the equation $f$ which is given by
\begin{equation} \label{eqn:add8}
    \mu \left( p^n_{h_1} p^n_{A_1} + p^n_{h_2} p^n_{A_2} + \sigma^2_r \lambda^n_{Bh}\right)=
   \sum^2_{i=1} \frac{\sigma^2_i \lambda^n_{Bg_i} p^n_{h_{\bar i}} p^n_g p^n_{A_{\bar i}}}
    {\left[\sigma^2_i \lambda^n_{Bg_i} + p^n_{A_r} \left(\sigma^2_r \lambda^n_{Bh} p^n_g + p^n_g p^n_{h_{\bar i}} p^n_{A_{\bar i}}\right)\right]^2},
\end{equation}
and the variable $\mu$ should be chosen to satisfy
\begin{equation} \nonumber
   \sum^N_{n=1} p^n_{A_r} \left( p^n_{h_1} p^n_{A_1} + p^n_{h_2} p^n_{A_2} + \sigma^2_r \lambda^n_{Bh}\right) =\tau_r.
\end{equation}

\subsubsection{Sub-problem~2}
For given $p^n_{A_r}$, $\forall n$, we obtain $p^n_{A_1}$ and $p^n_{A_2}$  by solving the optimization problem as follows:
\begin{eqnarray} \label{eqn:subprobelem2}
    && \min_{p^n_{A_1}, p^n_{A_2}, \forall n }  J^u_1 + J^u_2 \\ \nonumber
    {s.t.} && ~\sum^N_{n=1} p^n_{A_1}  \leq \tau_1, ~
    \sum^N_{n=1} p^n_{A_2}  \leq \tau_2,~ p^n_{A_1} \geq 0,~p^n_{A_2} \geq 0,~ \forall n \\ \nonumber
    && \sum^N_{n=1}  p^n_{A_r} \left ( p^n_{h_1} p^n_{A_1} + p^n_{h_2} p^n_{A_2} + \sigma^2_r {\lambda^n_{Bh}} \right ) \leq \tau_r
\end{eqnarray}
Also by verifying
\begin{equation} \label{eqn:partialA1} \nonumber
    \frac{\partial^2 J^u}{{\partial p^n_{A_i}}^2} =
    \frac{2\left(\sigma^2_{\bar i} \lambda^n_{Bg_{\bar i}} + \sigma^2_r \lambda^n_{Bh} p^n_g p^n_{A_r}\right)\left(p^n_{h_i} p^n_g p^n_{A_r}\right)^2}
    {\left[\sigma^2_{\bar i} \lambda^n_{Bg_{\bar i}} + \sigma^2_r \lambda^n_{Bh} p^n_g p^n_{A_r}  +p^n_{h_i} p^n_g p^n_{A_r}  p^n_{A_i}\right]^3} >0,~i=1,2
\end{equation}
the sub-problem \eqref{eqn:subprobelem2} is still convex. However, a closed-form solution to this problem is generally not available. Some standard numerical methods, such as interior-point method, can be used to get the optimum solution.

The solutions in \textit{Sub-problem 1} and \textit{Sub-problem 2} show that ${\bf P}_{A_r}$, ${\bf P}_{A_1}$ and ${\bf P}_{A_2}$ are tightly coupled. Thus, we apply an iterative approach to find the final solution.
As verified by our simulation, the algorithm converges in only a few iterations.
After obtaining ${\bf \Lambda}_{A_1}$, ${\bf \Lambda}_{A_2}$ and ${\bf \Lambda}_{A_r}$ from the square root of ${\bf P}_{A_1}$, ${\bf P}_{A_2}$ and ${\bf P}_{A_r}$, we substitute them into \eqref{eqn:A1} to get the precoders.

The overall algorithm is outlined as follows:

\vspace{-0.3cm}
\hrulefill
\par
{\footnotesize
\textbf{Algorithm 2} (Channel parallelization based precoding)
\begin{itemize}
\item \textbf{Decompose} the channel pairs $\{{\bf H}_1, {\bf H}_2\}$ and $\{{\bf G}_1, {\bf G}_2\}$ by using
\eqref{eqn:GSVDHnew} and \eqref{eqn:GSVDG}, respectively, to get ${\bf \Lambda}_{h_1}$, ${\bf \Lambda}_{h_2}$, ${\bf \Lambda}_{g}$, ${\bf B}_{g_1}$,
${\bf B}_{g_2}$ and ${\bf B}_h$.
\item \textbf{Repeat}
\begin{itemize}
\item Update the relay power allocation $p^n_{A_r}$ using \eqref{eqn:KKT8} to get ${\bf \Lambda}_{A_r}$;
\item Update the source power allocation $p^n_{A_1}$ and $p^n_{A_2}$ by solving \eqref{eqn:subprobelem2}
to get ${\bf \Lambda}_{A_1}$ and ${\bf \Lambda}_{A_2}$;
\end{itemize}
\item \textbf{Until} termination criterion is satisfied.
\item \textbf{Substitute} the solved ${\bf \Lambda}_{A_1}$, ${\bf \Lambda}_{A_2}$ and ${\bf \Lambda}_{A_r}$ into \eqref{eqn:A1} to get the precoders ${\bf A}_1$, ${\bf A}_2$ and ${\bf A}_r$.
\end{itemize}}
\vspace{-0.5cm}
\hrulefill

\section{Source-antenna-selection based Precoding for Single Data Stream}
In this section, we consider the precoding design for
the special case where only a single data stream is transmitted from each source.
The iterative approach proposed in Section III can be applied directly, except that the source precoding matrices reduce to beamforming vectors.
In what follows, we introduce a new
precoding strategy
based on antenna selection at two sources.
Antenna selection can be viewed as a special case of beamforming. In general, it is computationally less complex and requires lower feedback overhead. This motivates us to consider the source antenna selection while using precoding at the relay node only.

For single-data-stream transmission, the received signals ${\bf y}_i$ given in \eqref{eqn:source1receivedsignalnewnew} at each destination node is simplified as
\begin{equation}\label{SecV:y1}\nonumber
     {\bf y}_i =\sqrt{\tau_{\bar i}}{\bf G}_i {\bf A}_r {\bf h}_{{\bar i}n}  s_{\bar i}
     + {\bf G}_i {\bf A}_r {\bf n}_r + {\bf n}_i,~i=1,2
\end{equation}
where ${\bf h}_{{\bar i}n}$ is the selected forward channel vectors for $S_{\bar i}$ in the MAC phase. After decoding by ${\bf w}_i$, the corresponding MSE at $S_i$ is denoted as
\begin{equation}\label{SecV:J1}\nonumber
     J_i={\bf w}^H_i {\bf G}_i {\bf A}_r {\bf R}_{x_{\bar i}} {\bf A}^H_r {\bf G}^H_i {\bf w}_i
     - \sqrt{\tau_{\bar i}} {\bf w}^H_i {\bf G}_i {\bf A}_r {\bf h}_{{\bar i}n} - \sqrt{\tau_{\bar i}} {\bf h}^H_{{\bar i}n}{\bf A}^H_r {\bf G}^H_i {\bf w}_i
     + \sigma^2_i {\bf w}^H_i {\bf w}_i +1,~i=1,2
\end{equation}
where ${\bf R}_{x_i}=\tau_i {\bf h}_{in}{\bf h}^H_{in}+\sigma^2_r {\bf I}_{M}$. Thus, for a given selected antenna pair $\{{\bf h}_{1n},{\bf h}_{2m}\}$, the optimization problem is formulated as
\begin{equation}\label{SecV:OptPro}\nonumber
\begin{split}
     &~~~~~~ \min_{{\bf A}_r, {\bf w}_1, {\bf w}_2} J_1 +J_2\\
     &s.t. ~~{\rm Tr}\left\{ {\bf A}_r \left( \tau_1 {\bf h}_{1n}{\bf h}^H_{1n}+ \tau_2 {\bf h}_{2m}{\bf h}^H_{2m}
     + \sigma^2_r {\bf I}_{M} \right){\bf A}^H_r \right\}\leq \tau_r
\end{split}
\end{equation}
Next, we take two steps to solve ${\bf w}_1$, ${\bf w}_2$ and ${\bf A}_r$, respectively. First, for fixed ${\bf A}_r$, the optimal ${\bf w}_i$ is denoted as
\begin{equation}\label{SecV:Optw1}
    {\bf w}^{opt}_i = \left[{\bf G}_i {\bf A}_r {\bf R}_{x_{\bar i}} {\bf A}^H_r {\bf G}^H_i + \sigma^2_i {\bf I}_{M} \right]^{-1}
    {\bf G}_i {\bf A}_r {\bf h}_{{\bar i}n},~i=1,2.
\end{equation}
Subsequently, for fixed ${\bf w}_1$ and ${\bf w}_2$, we obtain the optimal ${\bf A}_r$ as
\begin{equation}\label{SecV:OptAr}
    {\bf A}^{opt}_r=
mat \left\{ {\bf R}^T_{x_2}\otimes \left( {\bf G}^H_1 {\bf w}_1 {\bf w}^H_1 {\bf G}_1 \right)
                       +{\bf R}^T_{x_1}\otimes \left( {\bf G}^H_2 {\bf w}_2 {\bf w}^H_2 {\bf G}_2 \right)+ \mu {\bf R}^T_x \otimes {\bf I}_{M} \right\}^{-1}vec\left\{{\bf M} \right\},
\end{equation}
where ${\bf R}_x=\tau_1 {\bf h}_{1n}{\bf h}^H_{1n}+\tau_2 {\bf h}_{2m}{\bf h}^H_{2m}+\sigma^2_r {\bf I}_{M}$, ${\bf M}=\sqrt{\tau_1}{\bf G}^H_2 {\bf w}_2 {\bf h}^H_{1n}+\sqrt{\tau_2}{\bf G}^H_1 {\bf w}_1 {\bf h}^H_{2m}$ and $\mu \in [0,\sqrt{\frac{{\rm Tr}\left\{{\bf M} {\bf R}^{-1}_x {\bf M}^H\right\}} {\tau_r}} ]$ is chosen to satisfy the KKT conditions.
The derivation is similar to the steps derived in Section III, and hence omitted for brevity.
In summary, we outline the algorithm as follows:

\vspace{-0.3cm}
\hrulefill
\par
{\footnotesize
\textbf{Algorithm 3} (Source antenna selection (SAS)-based precoding)
\begin{itemize}
\item \textbf{For} each source antenna pair $\{{\bf h}_{1n},{\bf h}_{2m}\}$, $\forall n, m$
\begin{itemize}
\item \textbf{Initialize} ${\bf A}_r$ randomly or as $\sqrt{\frac{\tau_r}{{\rm Tr}\left\{{\bf R}_x \right\}}}{\bf I}_{M}$ with ${\bf R}_x= \tau_1 {\bf h}_{1n}{\bf h}^H_{1n}+ \tau_2 {\bf h}_{2m}{\bf h}^H_{2m}
     + \sigma^2_r {\bf I}_{M}$
\item \textbf{Repeat}
\begin{itemize}
\item Update the decoding vector by \eqref{SecV:Optw1} for fixed ${\bf A}_r$;
\item Update the relay precoder by \eqref{SecV:OptAr} for fixed ${\bf w}_1$ and ${\bf w}_2$;
\end{itemize}
\item \textbf{Until} termination criterion is satisfied.
\end{itemize}
\item \textbf{End} choose the source antenna pair and the corresponding ${\bf w}_1$, ${\bf w}_2$ and ${\bf A}_r$ that lead to the minimal Total-MSE $J_1+J_2$.
\end{itemize}}
\vspace{-0.5cm}
\hrulefill

\textit{Remark 2}: Compared with the three-step iterative precoding algorithm, Algorithm 1, the SAS-based precoding algorithm, Algorithm 3, only needs two steps in each iteration. Additionally, the closed-form solution can be employed in each iteration. Thus,
no advanced software package is needed here.

\section{Simulation results and Discussions}
In this section, we present some simulation examples to evaluate the proposed precoding designs. The channel is set to be Rayleigh fading, i.e., the elements of each channel matrix are complex Gaussian random variables with zero mean and unit variance. For simplicity, we consider the reciprocal channel where ${\bf G}_1={\bf H}^T_1$ and ${\bf G}_2={\bf H}^T_2$ (our algorithm is suitable for the general case where ${\bf G}_i$ are ${\bf H}_i$ are independent). The noise powers at two destinations are set to be equal to each other, i.e., $\sigma^2_1=\sigma^2_2=\sigma^2$. The average signal-to-noise ratios (SNRs) for the MAC phase and BC phase are defined as $\rho_1=\frac{\tau_1}{\sigma^2_r}$, ${\rho_2}=\frac{\tau_2}{\sigma^2_r}$ and $\rho_r=\frac{\tau_r}{\sigma^2}$, respectively.
 The average bit error rate (BER) using quadrature phase-shift
keying (QPSK) modulation is simulated.

\subsection{Convergence and Robustness of the Proposed Iterative Algorithm}
\begin{figure}[tbhp]
\begin{centering}
\includegraphics[scale=0.80]{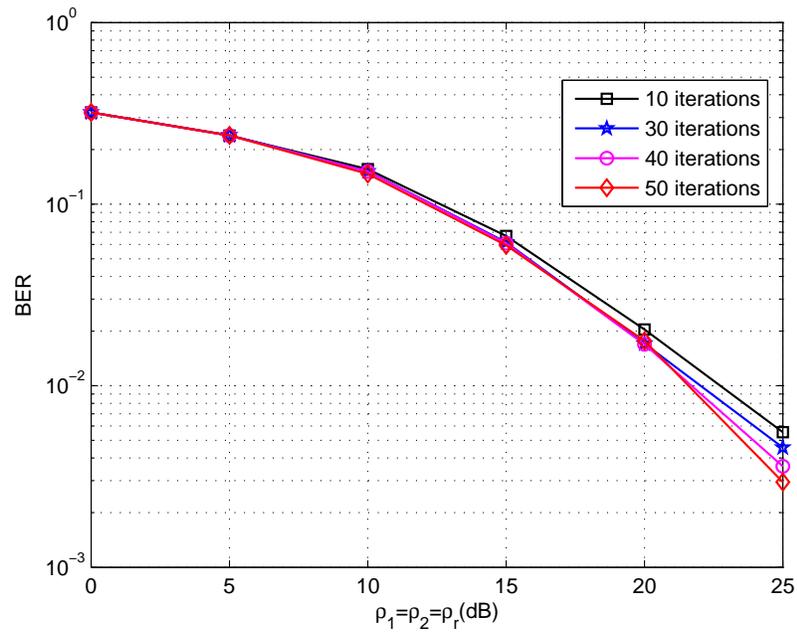}
\vspace{-0.1cm}
\caption{Convergence behavior of the proposed iterative precoding algorithm.} \label{fig:iteTotalMSE}
\end{centering}
\vspace{-0.3cm}
\end{figure}

Fig.~\ref{fig:iteTotalMSE} illustrates the convergence behavior of the iterative algorithm presented in Section III as the function of SNR at $N=M=2$. It is found that, in the low SNR regime, the iterative algorithm converges within $10$ iterations.
With medium SNR, it converges after about $30$ iterations. While in the high SNR regime, $50$ iterations are always enough.

\begin{figure}[tbhp]
\begin{centering}
\includegraphics[scale=0.80]{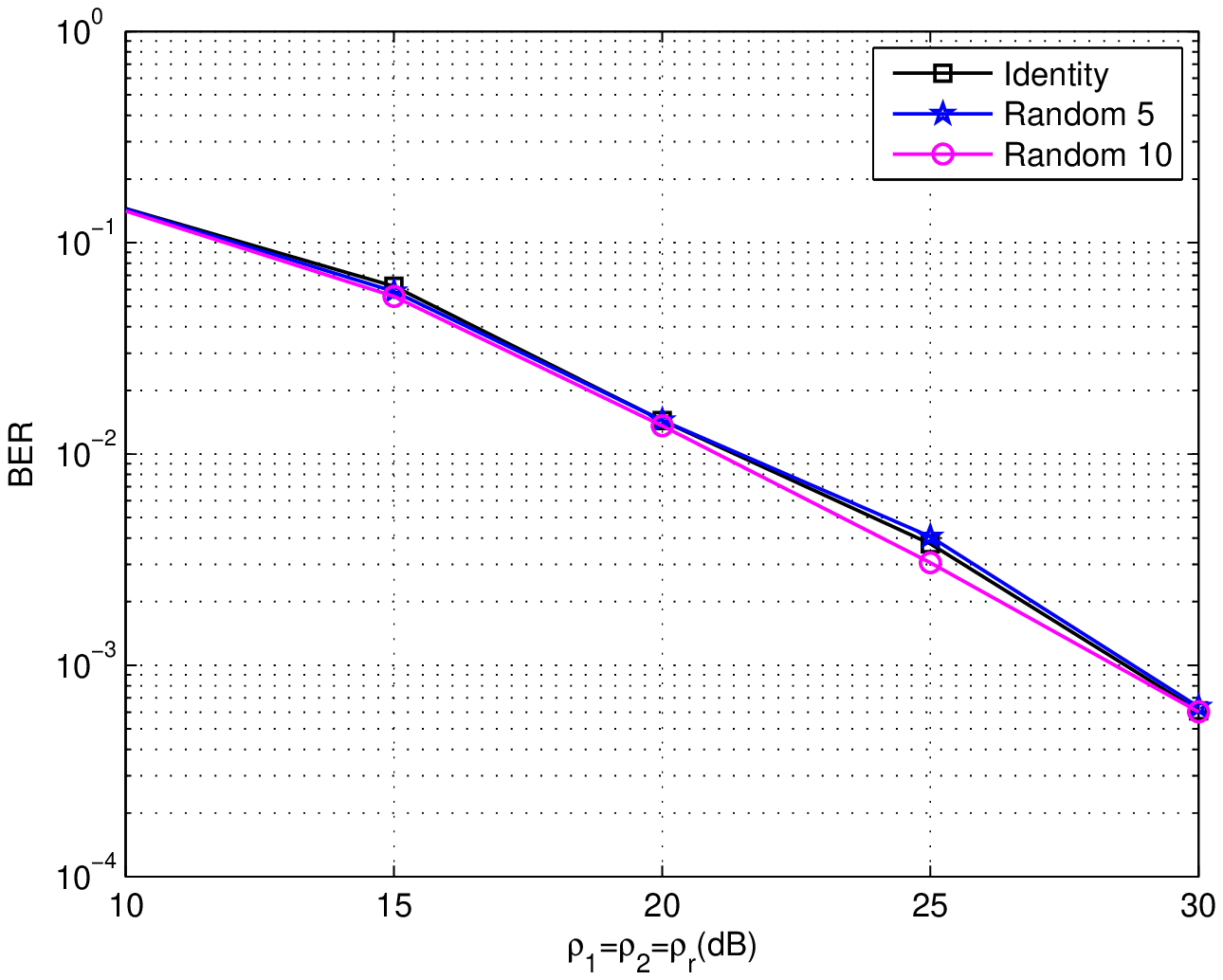}
\vspace{-0.1cm}
\caption{Performance comparison of iterative algorithm with different initialization points at $N=M=2$.} \label{fig:Random}
\end{centering}
\vspace{-0.3cm}
\end{figure}

\begin{figure}[tbhp]
\begin{centering}
\includegraphics[scale=0.80]{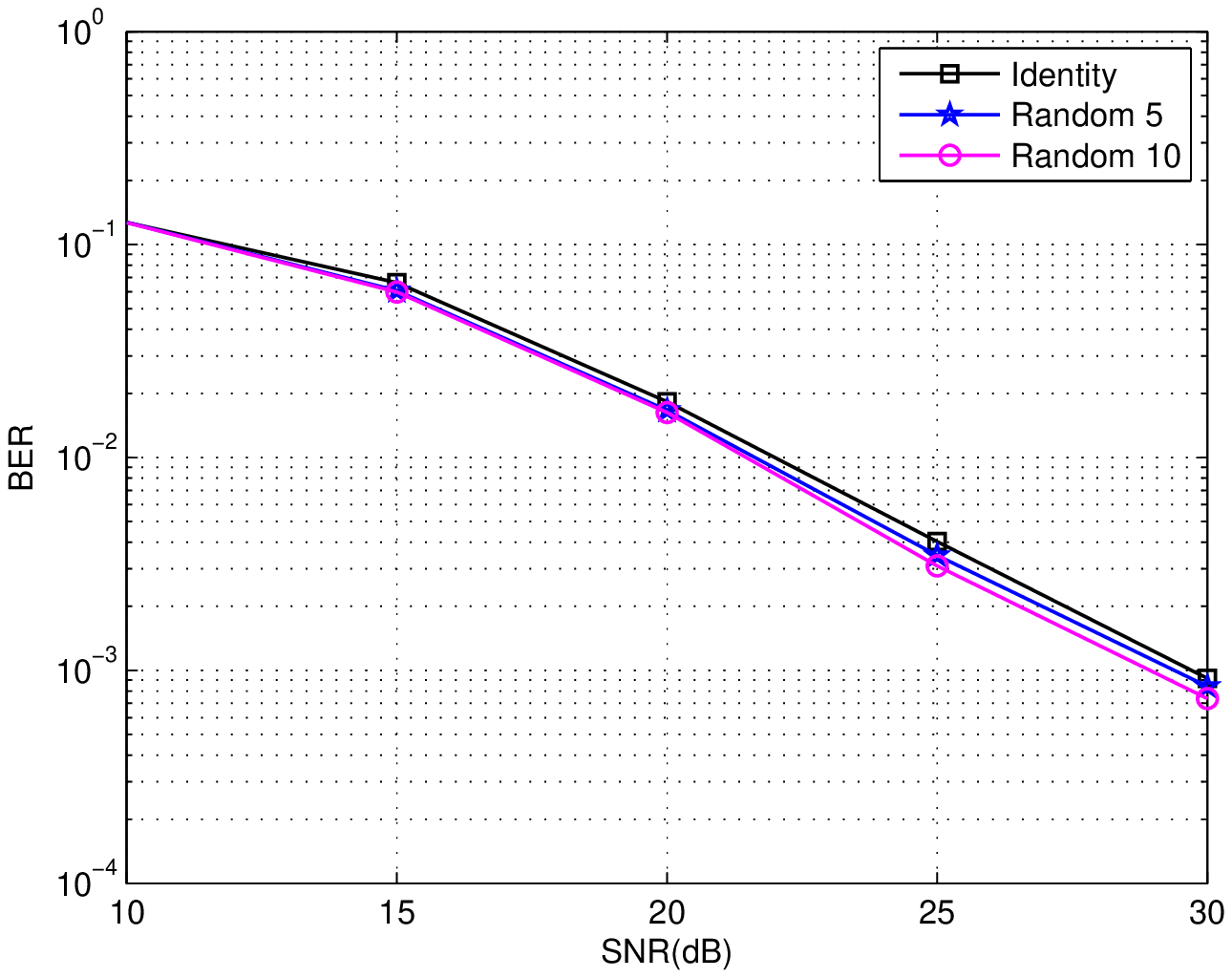}
\vspace{-0.1cm}
\caption{Performance comparison of iterative algorithm with different initialization points at $N=M=4$.} \label{fig:Random444}
\end{centering}
\vspace{-0.3cm}
\end{figure}

Since the proposed iterative precoding algorithm only finds the local optimal solution due to non-convexity of the primal problem, different initialization points
may result in different
convergent solutions. Fig.~\ref{fig:Random} and Fig.~\ref{fig:Random444} show performance comparison with different initialization points at $N=M=2$ and $N=M=4$, respectively. Here, ``Identity" means that the algorithm is initialized by the identity matrix, while ``Random $N$" means that $N$ randomly generated initialization points are tried and the one with the best performance is finally chosen. We observe
that the BER performance gain by choosing the best out of different initialization points is minimal.
We thus conclude that the proposed iterative precoding algorithm is robust to the initialization points and hence near optimal. For the rest of the simulation, the ``Identity" initialization point is adopted unless specified otherwise.

\subsection{Performance Comparison for Multi-data-stream Transmission}
\begin{figure}[tbhp]
\begin{centering}
\includegraphics[scale=0.80]{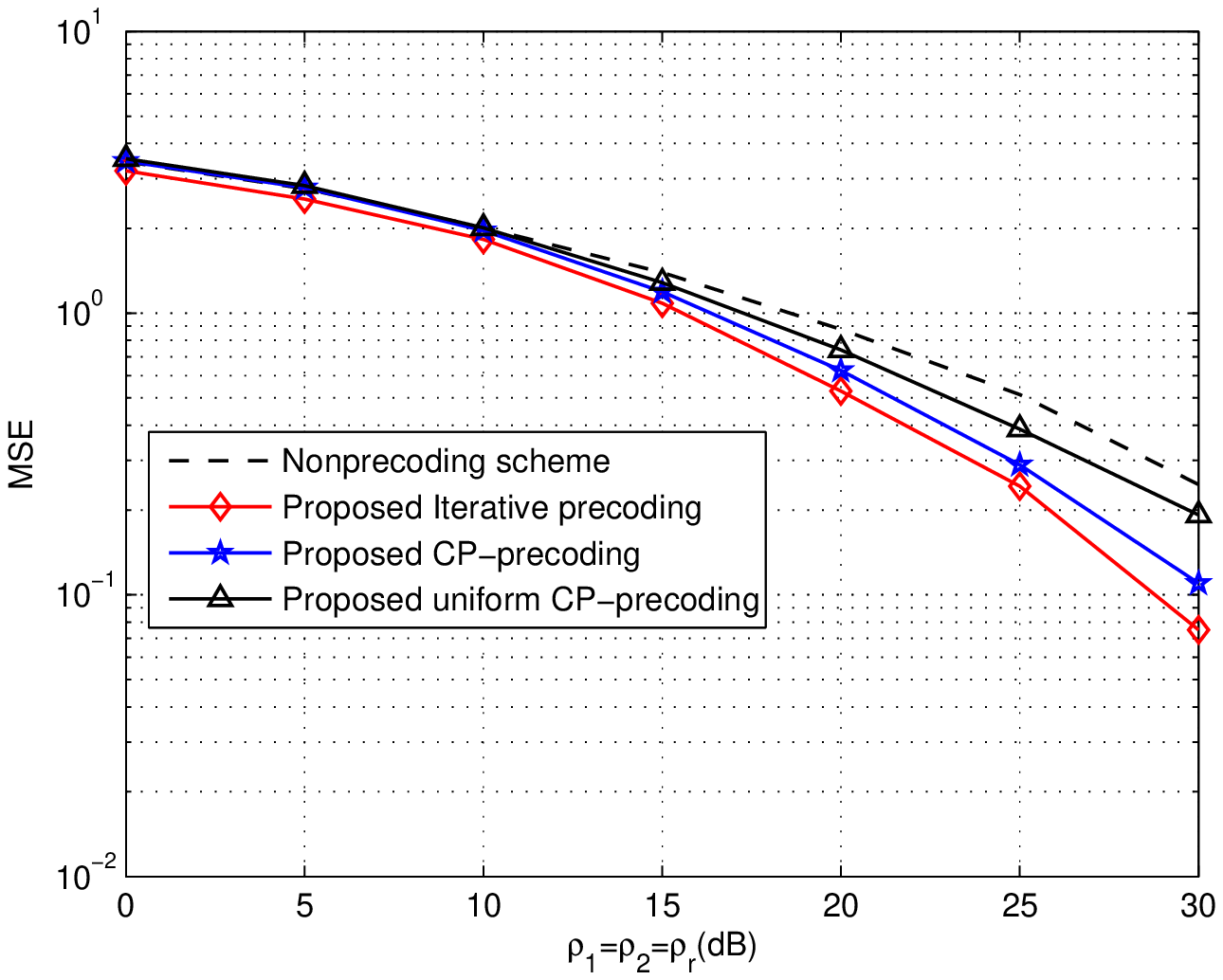}
\vspace{-0.1cm}
\caption{The MSE performance comparison with $\rho_1=\rho_2=\rho_r$ at $N=M=2$.} \label{fig:AllCopmMSE}
\end{centering}
\vspace{-0.3cm}
\end{figure}

\begin{figure}[tbhp]
\begin{centering}
\includegraphics[scale=0.80]{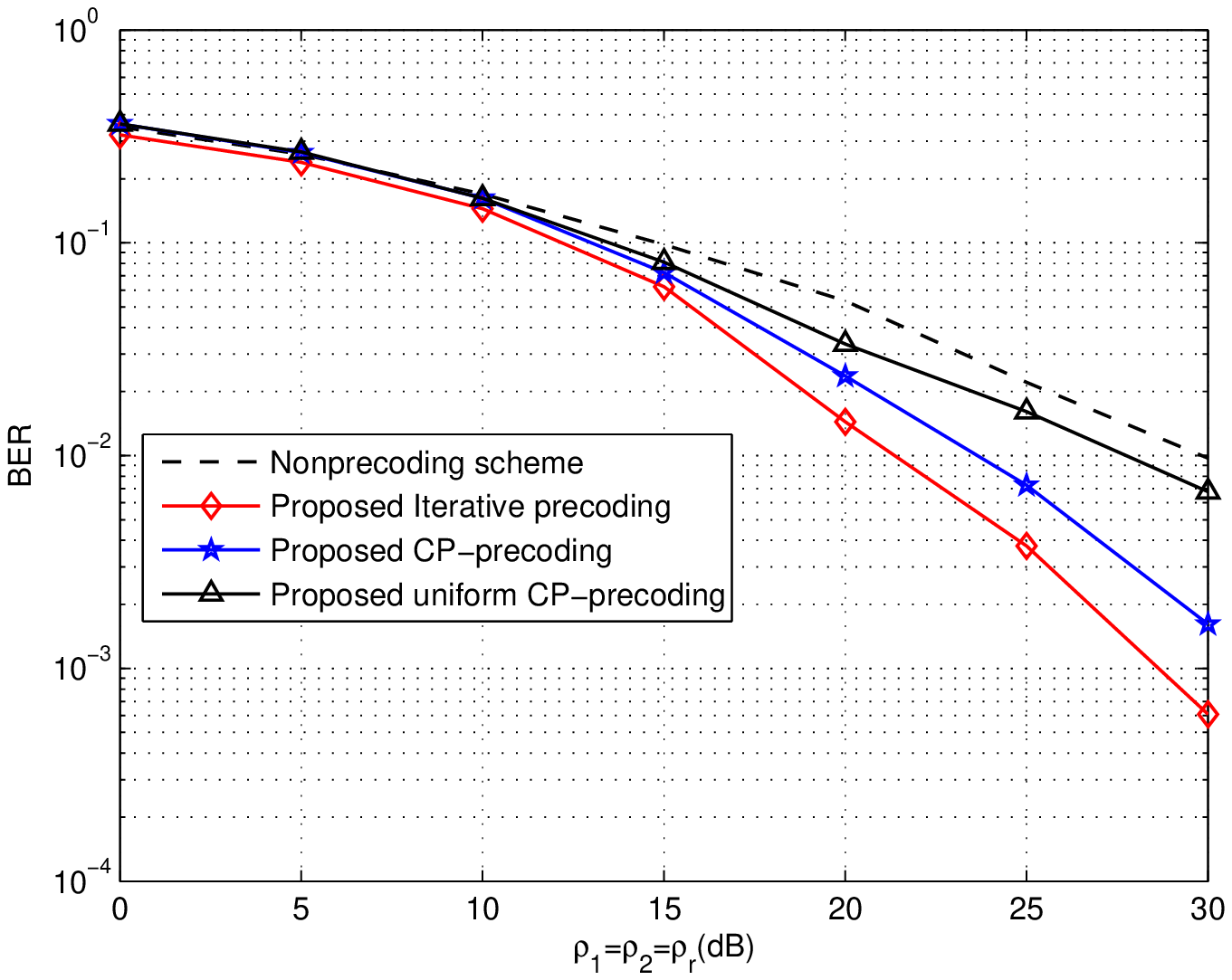}
\vspace{-0.1cm}
\caption{The BER performance comparison with $\rho_1=\rho_2=\rho_r$ at $N=M=2$.} \label{fig:AllCopmBER}
\end{centering}
\vspace{-0.3cm}
\end{figure}

In Fig.~\ref{fig:AllCopmMSE} and Fig.~\ref{fig:AllCopmBER}, we show the MSE and BER performance comparison of the proposed iterative precoding design and the channel-parallelization based precoding design (CP-precoding)
as the function of $\rho_1=\rho_2=\rho_r$ at $N=M=2$. For comparison,
the CP-precoding design with uniform power allocation (uniform CP-precoding), i.e.,
equal power distribution among
all data streams, is also simulated. We find that with both the
iterative precoding and the CP-precoding, the system BER decreases considerably when SNR increases.
This demonstrates the effectiveness of the proposed precoding designs.
We also find that
the uniform CP-precoding only achieves marginal gain over the non-precoding case. This is due to the fact that
uniform power allocation can lead to unfair channel gain distribution among the data streams, and the system BER performance is dominated by the poorest sub-channel.
We thus conclude that it is essential to optimize the power allocation among data streams for the channel-parallelization based precoding design.
From Fig.~\ref{fig:AllCopmMSE} and Fig.~\ref{fig:AllCopmBER},
it is observed that the iterative precoding
designs exhibits the best performance among all the proposed precoding designs. We attribute the performance improvement to not enforcing any structure on the
precoders.

\begin{figure}[tbhp]
\begin{centering}
\includegraphics[scale=0.80]{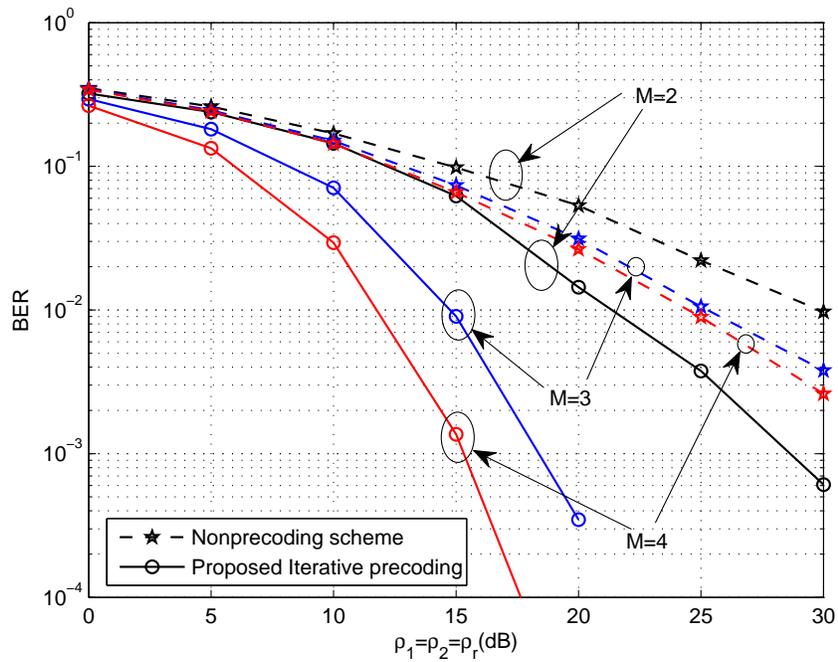}
\vspace{-0.1cm}
\caption{The BER performance comparison for different relay antenna number.} \label{fig:ber_changeNr}
\end{centering}
\vspace{-0.3cm}
\end{figure}

Fig.~\ref{fig:ber_changeNr} illustrates the BER performance comparison at different relay antenna number $M$ when the source antenna number is fixed at $N=2$. We find that increasing the relay antennas significantly enhances the BER performance thanks to the increased diversity gain.
Moreover, the gain of the proposed precoding scheme over
the non-precoding scheme
increases dramatically as the number of relay antennas increases.
It further implies that when the relay node has more antennas than the source nodes, conducting the precoding is more beneficial.

\begin{figure}[tbhp]
\begin{centering}
\includegraphics[scale=0.80]{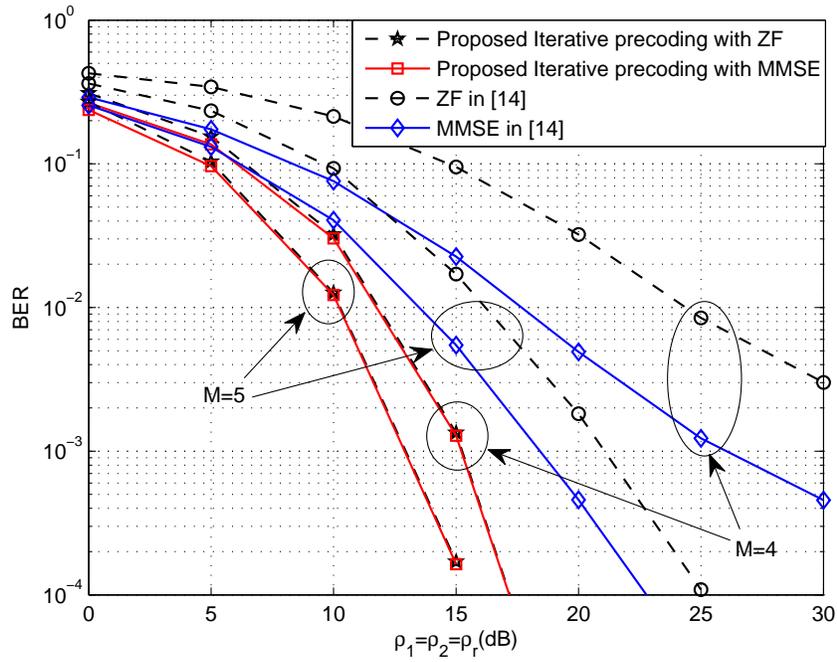}
\vspace{-0.1cm}
\caption{The BER performance comparison with \cite{Timo2008}.} \label{fig:ber_comparison}
\end{centering}
\vspace{-0.3cm}
\end{figure}

Finally, the performance comparison between the proposed iterative joint source/relay precoding and the relay precoding scheme in \cite{Timo2008} is depicted in Fig.~\ref{fig:ber_comparison} at $N=2$. Since the antenna configuration in \cite{Timo2008} should satisfy the condition $M\geq 2N$, we choose $M=4$ and $5$ in the simulation. It is shown that, by applying either MMSE or ZF receiver, the proposed joint source/relay precoding significantly outperforms the scheme in \cite{Timo2008} where precoding is applied at the relay only. This implies that
in two-way relay systems, precoding at the source nodes is very helpful in improving the system performance.
It is also found that both MMSE and ZF receivers obtain almost the same performance for the proposed precoding algorithm.

\subsection{Performance Comparison for Single-data-stream Transmission}
\begin{figure}[tbhp]
\begin{centering}
\includegraphics[scale=0.80]{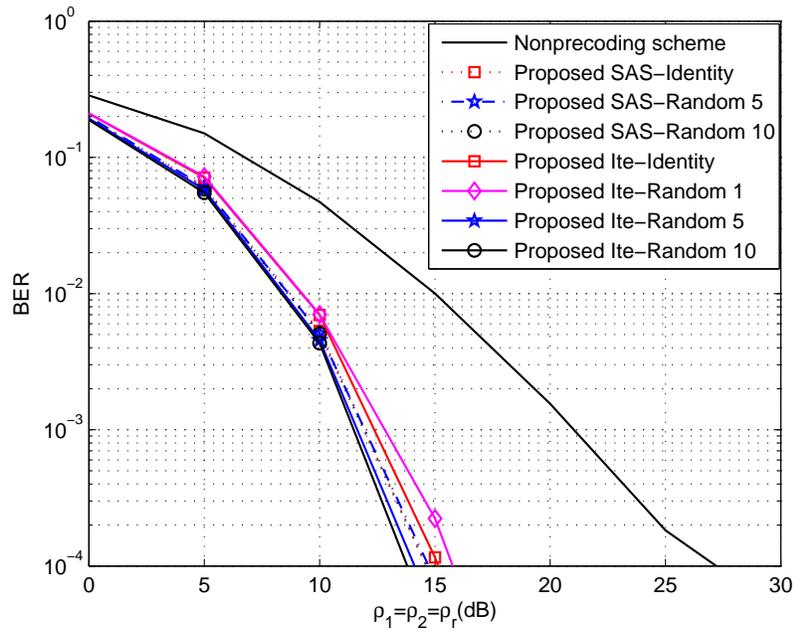}
\vspace{-0.1cm}
\caption{The BER performance comparison with $\rho_1=\rho_2=\rho_r$ at $N=M=2$ for single-data-stream transimssion.} \label{fig:NewSingle}
\end{centering}
\vspace{-0.3cm}
\end{figure}

In Fig.~\ref{fig:NewSingle}, we show the BER performance
for single-data-steam
transmission.
Here, the proposed iterative precoding (proposed ite-precoding) and the source-antenna-selection based precoding (proposed SAS-precoding) are simulated.
We find that the performance gained through precoding is more significant for the single-data-stream transmission than for the multi-data-stream transmission. This is because there is no interference from other data streams.
In addition,
with the ``Identity" initialization point, the SAS-precoding
almost has the same performance as ``Random $5$" and ``Random $10$" cases
\footnote{ It implies that the ``Identity" relay precoding matrix is usually a good initialization point as in the multi-data-stream case.},
and it
outperforms the ite-precoding method with both ``Identity"\footnote{For ite-precoding method, only the relay precoder is the matrix, while two source precoders is actually vectors. Here, with slight confused using of the notation, ``Identity" source precoder means the vector with equal entries.}  and ``Random $1$" initialization point although it needs lower feedback overhead. The reason is that the optimal beamforming vector at each source cannot be obtained due to the non-convexity nature of the joint optimization problem, while by exhaustively searching the most suitable source antenna pair, the SAS-precoding design can achieve better performance.
However, as the number of randomly generated initialization points increases, the ite-precoding design starts to outperform the SAS-precoding design
\footnote{Note here it is different from the multi-data-stream iterative precoding, we find that the ``Identity" source precoding vector is not a good initialization point.},
as the ite-precoding design is approaching the optimal solution.
Moreover, it is shown that the ``Random $5$" ite-precoding design scheme and the ``Random $10$" ite-precoding design scheme almost obtain the same performance. However, such optimal approaching solution has substantially higher computational complexity and may not be practical for implementation.

\section{Conclusions}
In this paper, we studied the joint source/realy precoding design for AF MIMO two-way relay systems based on the MSE criterion.
An iterative method was first proposed to obtain the local optimal solutions for the Total-MSE minimization.
Then, for the scenario in which all nodes are equipped with the same number of antennas, we proposed a channel-parallelization based precoding design algorithm
to parallelize the channels in both MAC and BC phases. By doing so, the joint precoder design is reduced to a simple power allocation problem.
It was shown that
the iterative precoding design
outperforms the channel-parallelization based precoding design since no structure constraint is enforced on the precoders. Although the channel-parallelization method obtains degraded performance, it on the other hand reduces the computational complexity.
When single data stream is transmitted from each source,
the precoding at source nodes can be replaced by the antenna selection.
By this way, the system feedback overhead is reduced and no advanced software package is needed.
Simulation results showed that
all the proposed precoding designs are effective compared with conventional schemes.

\appendices
\section{Proof of lemma 1}
\label{prof_lemma1}
We first show that the objective function in \eqref{Optimization2} is a convex function.
Since the sum of two convex functions is still a convex function,
the convexity of $J_{r_1}+J_{r_2}$ can be verified by showing that $J_{r_1}$ and $J_{r_2}$ are both convex.
We take $J_{r_1}$ as the example to illustrate the proof and the extension to $J_{r_2}$ is straightforward.
For notation simplicity, we define
${\bf R}_1={\bf G}^H_1 {\bf W}^H_1 {\bf W}_1 {\bf G}_1$,
${\bf R}_2={\bf H}_2 {\bf A}_2 {\bf W}_1 {\bf G}_1$ and
$a={\rm Tr}\left\{\sigma^2_1 {\bf W}_1 {\bf W}^H_1 +{\bf I}_N \right\}$.
By applying matrix manipulations in \cite[Eq.1.10.62, Eq.1.10.64]{xiandazhang2004},
$J_{r_1}$ can be reformulated as
\begin{equation}\label{NewJr1}\nonumber
    J_{r_1}={\bf a}^H_r \left({\bf R}_{x_2}^T \otimes {\bf R}_1 \right ) {\bf a}_r
    + vec({\bf R}^T_2)^T  {\bf a}_r
    + {\bf a}^H_r vec({\bf R}^H_2)
    + a,
\end{equation}
where ${\bf a}_r=vec({\bf A}_r)$.
Based on the vector differential rule in \cite{Hjorungnes2007}, four Hessian matrices as defined in \cite{Hjorungnes2007a} are derived as
\begin{equation}\label{Hessian1}\nonumber
    {\cal H}_{{\bf a}^{*}_r,{\bf a}_r}J_{r_1}=({\bf R}_{x_2}^T \otimes {\bf R}_1)^T,
    ~~{\cal H}_{{\bf a}_r,{\bf a}^{*}_r}J_{r_1}={\bf R}_{x_2}^T \otimes {\bf R}_1,
    ~~{\cal H}_{{\bf a}_r,{\bf a}_r}J_{r_1}={\bf 0},
    ~~{\cal H}_{{\bf a}^{*}_r,{\bf a}^{*}_r}J_{r_1}={\bf 0}.
\end{equation}
In order to show the convexity of $J_{r_1}$, the following block matrix should be positive semidefinite
\begin{eqnarray}\label{BlockHessian}\nonumber
    {\cal H} (J_{r_1})=\left [ \begin{array}{cc}
                      {\bf R}_{x_2}^T \otimes {\bf R}_1 & {\bf 0} \\
                      {\bf 0} & \left({\bf R}_{x_2}^T \otimes {\bf R}_1\right)^T
                      \end{array}\right].
\end{eqnarray}
Before confirming the positive semidefinition of ${\cal H} (J_{r_1})$, we introduce the following lemma.

\textbf{Lemma A}:
The Kronecker product of any two positive semidefinite matrices is also positive semidefinite.
\begin{proof}
Let ${\bf Z}_1$ and ${\bf Z}_2$ be any two positive semidefinite  matrices. We can decompose them into ${\bf Z}_1={\bf Z}^{\frac{1}{2}}_1{\bf Z}^{\frac{1}{2}}_1$ and ${\bf Z}_2={\bf Z}^{\frac{1}{2}}_2{\bf Z}^{\frac{1}{2}}_2$ where ${\bf Z}^{\frac{1}{2}}_1$ and ${\bf Z}^{\frac{1}{2}}_2$ are also both positive semidefinite matrices. Applying the rule ${\bf A}{\bf B}\otimes {\bf C}{\bf D}=({\bf A}\otimes {\bf C})({\bf B}\otimes {\bf D})$, we have
\begin{equation}\label{lemma2-1}\nonumber
   {\bf Z}={\bf Z}_1 \otimes {\bf Z}_2
          =\left({\bf Z}^{\frac{1}{2}}_1{\bf Z}^{\frac{1}{2}}_1\right) \otimes \left({\bf Z}^{\frac{1}{2}}_2{\bf Z}^{\frac{1}{2}}_2 \right)
          =\left({\bf Z}^{\frac{1}{2}}_1\otimes {\bf Z}^{\frac{1}{2}}_2\right)\left({\bf Z}^{\frac{1}{2}}_1\otimes {\bf Z}^{\frac{1}{2}}_2 \right).
\end{equation}
Since ${\bf Z}^{\frac{1}{2}}_1\otimes {\bf Z}^{\frac{1}{2}}_2$ is Hermitian, we conclude that matrix ${\bf Z}$ is positive semidefinite.
\end{proof}

By applying Lemma A, we derive that the matrix ${\bf R}_{x_2}^T \otimes {\bf R}_1$
is positive semidefinite since both ${\bf R}_{x_2}^T $ and ${\bf R}_1$ are positive semidefinite.
Then, ${\cal H} (J_{r_1})$ is positive semidefinite.
Hence, the convexity of $J_{r_1}$ is proven.
The same result holds for $J_{r_2}$.
Thus we conclude that the objective function $J_{r_1}+J_{r_2}$ is convex.

Next, we prove that the feasible set provided by ${\rm Tr} \left\{ {\bf A}_r {\bf R}_x {\bf A}^H_r \right\} \leq \tau_r$ is convex. This can be alternatively proven by checking the convexity of the function
    $f={\rm Tr} \left\{ {\bf A}_r {\bf R}_x {\bf A}^H_r \right\}$ \cite{Boyd2004}.
Similar to the previous manipulation, $f$ can be reexpressed as
    $f={\bf a}^H_r ({\bf R}^T_x \otimes {\bf I}_{M}) {\bf a}_r$.
In addition, the corresponding four Hessian matrices are derived as
\begin{equation}\label{Hessian2}\nonumber
    {\cal H}_{{\bf a}^{*}_r,{\bf a}_r}f=({\bf R}_{x_2}^T \otimes {\bf I}_{M})^T,
    ~~{\cal H}_{{\bf a}_r,{\bf a}^{*}_r}f={\bf R}_{x_2}^T \otimes {\bf I}_{M}\overleftarrow{},
    ~~{\cal H}_{{\bf a}_r,{\bf a}_r}f={\bf 0},
    ~~{\cal H}_{{\bf a}^{*}_r,{\bf a}^{*}_r}f={\bf 0}.
\end{equation}
Applying Lemma A, we can also show that the block matrix ${\cal H} (J_{f})$
is positive semidefinite. Thus, we derive that the feasible set in \eqref{Optimization2} is convex.
Since both the objective function and the feasible set are convex, the optimization problem \eqref{Optimization2} is a convex problem.

\section{Proof of lemma 2}
\label{prof_lemma2}
Since there exists an inverse operator outside the Lagrangian multiplier $\lambda$, it is easy to verify that $g$ decreases with $\lambda$. Next we mainly focus on deriving the upper bound of $\lambda$.
To this end,
we first assume that ${\bf R}_r$ can be divided into two parts as ${\bf R}_r= {\bf Q}_1+{\bf Q}_2$, and let
\begin{equation}\label{Q1}
    {\bf Q}_1= {\bf R}_{r_1} {\bf A}^{opt}_r {\bf R}_{x_2}+ {\bf R}_{r_2} {\bf A}^{opt}_r {\bf R}_{x_1},~
    {\bf Q}_2= {\lambda}^{opt} {\bf A}^{opt}_r {\bf R}_x,
\end{equation}
where ${\bf A}^{opt}_r$, $\lambda^{opt}$ are the optimal primal and dual solutions of \eqref{Optimization2}.
Applying \eqref{Q1}, we have
\begin{equation}\label{Q2Ar}
    {\bf A}^{opt}_r=\frac{1}{{\lambda}^{opt}}{\bf Q}_2 {\bf R}^{-1}_x.
\end{equation}
Substituting \eqref{Q2Ar} into the power constraint \eqref{KKTconditions3} to make the equality satisfied, it has
\begin{equation}\label{Q2power}\nonumber
    {\rm Tr}\left\{ {\bf A}^{opt}_r {\bf R}_x {{\bf A}^{opt}_r}^H \right\}
    ={\rm Tr}\left\{ \frac{1}{{\lambda^{opt}}^2}{\bf Q}_2 {\bf R}^{-1}_x {\bf R}_x {\bf R}^{-1}_x {\bf Q}^H_2 \right\}
    ={\rm Tr}\left\{ \frac{1}{{\lambda^{opt}}^2}{\bf Q}_2 {\bf R}^{-1}_x {\bf Q}^H_2\right\}=\tau_r.
\end{equation}
On the other hand, we have
\begin{equation}\label{Qpower}
\begin{split}
    {\rm Tr}\left\{ \frac{1}{{\lambda^{opt}}^2} {\bf R}_r {\bf R}^{-1}_x {\bf R}^H_r\right\}
    &= {\rm Tr}\left\{ \frac{1}{{\lambda^{opt}}^2} \left({\bf Q}_1+{\bf Q}_2\right) {\bf R}^{-1}_x \left({\bf Q}_1+{\bf Q}_2\right)^H \right\}\\
    &={\rm Tr}\left\{\frac{1}{{\lambda^{opt}}^2} {\bf Q}_1 {\bf R}^{-1}_x {\bf Q}^H_1 \right\}
    + {\rm Tr}\left\{\frac{1}{{\lambda^{opt}}^2} {\bf Q}_2 {\bf R}^{-1}_x {\bf Q}^H_2 \right\}\\
    &+{\rm Tr}\left\{\frac{1}{{\lambda^{opt}}^2} {\bf Q}_1 {\bf R}^{-1}_x {\bf Q}^H_2 \right\}
    +{\rm Tr}\left\{\frac{1}{{\lambda^{opt}}^2} {\bf Q}_2 {\bf R}^{-1}_x {\bf Q}^H_1 \right\}.
\end{split}
\end{equation}
Since if ${\bf Z}_1,~{\bf Z}_2$ are positive semidefinite, it has ${\rm Tr}\left\{{\bf Z}_1 {\bf Z}_2\right\}\geq 0$. We thus conclude that ${\rm Tr}\left\{\frac{1}{{\lambda^{opt}}^2} {\bf Q}_1 {\bf R}^{-1}_x {\bf Q}^H_1 \right\}$ in \eqref{Qpower} larger than or at least equal to zero. Next we prove ${\rm Tr}\left\{\frac{1}{{\lambda^{opt}}^2} {\bf Q}_1 {\bf R}^{-1}_x {\bf Q}^H_2 \right\} \geq0$. Based the definition in \eqref{Q1}, it has
\begin{equation}\label{Qpower1}
    {\rm Tr}\left\{ {\bf Q}_1 {{\bf A}^{opt}_r}^H \right\}
    = {\rm Tr}\left\{{\bf R}_{r_1} {\bf A}^{opt}_r {\bf R}_{x_2} {{\bf A}^{opt}_r}^H
    + {\bf R}_{r_2} {\bf A}^{opt}_r {\bf R}_{x_1} {{\bf A}^{opt}_r}^H \right\}\geq 0.
\end{equation}
Substituting \eqref{Q2Ar} into \eqref{Qpower1}, we obtain
\begin{equation}\label{Qpower2}\nonumber
    {\rm Tr}\left\{ {\bf Q}_1 {{\bf A}^{opt}_r}^H \right\}
    = {\rm Tr}\left\{ \frac{1}{\lambda^{opt}} {\bf Q}_1  {\bf R}^{-1}_x {\bf Q}^H_2\right\}.
\end{equation}
Thus, we conclude that ${\rm Tr}\left\{\frac{1}{{\lambda^{opt}}^2} {\bf Q}_1 {\bf R}^{-1}_x {\bf Q}^H_2 \right\} \geq0$
(the same for ${\rm Tr}\left\{\frac{1}{{\lambda^{opt}}^2} {\bf Q}_2 {\bf R}^{-1}_x {\bf Q}^H_1 \right\}$).
Since all terms in \eqref{Qpower} are larger than or at lease equal to zero, we conclude
\begin{equation}\label{upperboundtau1}\nonumber
    {\rm Tr}\left\{ \frac{1}{{\lambda^{opt}}^2} {\bf R}_r {\bf R}^{-1}_x {\bf R}^H_r\right\} \geq \tau_r.
\end{equation}
Thus, the proof of Lemma 2 is completed.

\section{Proof of Lemma 3}
\label{prof_Optimiztion3}
By using the rule of the trace operator ${\rm Tr}\left\{{\bf A} {\bf B} {\bf C} {\bf D}\right\}=\left(vec({\bf D})^T\right)^T \left({\bf C}^T \otimes {\bf A}\right) vec({\bf B})$ in \cite{xiandazhang2004}, $J_{s_i}$ can be reformulated as
\begin{equation}\label{Js1New1}
    J_{s_i}={\hat {\bf a}}^H_{\bar i} {\hat {\bf P}_i} {\hat {\bf a}}_{\bar i}
    - 2 {\Re}\left\{ {\hat {\bf b}^T}_i {\hat {\bf a}}_{\bar i} \right\} + {\rm Tr}\left\{ {\bf R}_{s_{i3}} \right\},~i=1,2
\end{equation}
where ${\hat {\bf P}_i}={\bf I}_N \otimes {\bf R}_{s_{i1}}$, ${\hat {\bf b}}_i=vec({\bf R}^T_{s_{i2}})$ and ${\hat {\bf a}}_i=vec({\bf A}_i)$.
Again it is known that $\hat {\bf P}_{i}$ is a positive semidefinite matrix from Lemma A. Thus, \eqref{Js1New1} can be transformed into
\begin{equation}\label{Js1New21}
    J_{s_i}=||{\hat {\bf P}^{\frac{1}{2}}_i}{\hat {\bf a}}_{\bar i}||^2_2
    - 2 {\Re}\left\{ {\hat {\bf b}^T}_i {\hat {\bf a}}_{\bar i} \right\} + {\rm Tr}\left\{ {\bf R}_{s_{i3}} \right\},~i=1,2.
\end{equation}
To further delete ${\Re}(\cdot)$ operator, we redefine ${\bf a}_i=\left[\Re\{{\hat {\bf a}^T}_i\}, \Im\{{\hat {\bf a}^T}_i\}\right]^T$, $i=1,2$ and transform \eqref{Js1New21} into
\begin{equation}\label{Js1New31}\nonumber
    J_{s_i}={\bf a}^T_{\bar i} {\bf P}_i {\bf a}_{\bar i}
    - 2  {\bf b}^T_i {\bf a}_{\bar i}  + {\rm Tr}\left\{ {\bf R}_{s_{i3}} \right\},~i=1,2
\end{equation}
where
${\bf P}_i={\tilde {\bf P}^T_i} {\tilde {\bf P}_i}$
with ${\tilde {\bf P}_i}=\left [ \begin{array}{cc}
                      \Re\left\{{\hat {\bf P}^{\frac{1}{2}}_i}\right\}
                      & -\Im\left\{{\hat {\bf P}^{\frac{1}{2}}_i}\right\}  \\
                      \Im\left\{{\hat {\bf P}^{\frac{1}{2}}_i}\right\} & \Re\left\{{\hat {\bf P}^{\frac{1}{2}}_i}\right\}
                      \end{array}\right]$,
${\bf b}_i=\left[\Re\{\hat{\bf b}^T_i\}, -\Im\{\hat{\bf b}^T_i\}\right]^T$, $i=1,2$. It is easy to verify that ${\bf P}_i$ is a positive semidefinite matrix. Similarly, for three power constraints, we have
    ${\rm Tr}\{ {\bf A}^H_i {\bf A}_i \}= {\bf a}^T_i {\hat {\bf Q}_i} {\bf a}_i$ with
    ${\hat{\bf Q}_i}={\bf I}_{2N^2 \times 2N^2}$, $i=1,2$,
    ${\rm Tr}\left\{ {\bf R}_{p_1}{\bf A}_1 {\bf A}^H_1
       + {\bf R}_{p_2} {\bf A}_2 {\bf A}^H_2 \right\}
       ={\bf a}^H_1 {\hat {\bf Q}_3} {\bf a}_1 +
       {\bf a}^H_2 {\hat {\bf Q}_4} {\bf a}_2$ with
${\hat {\bf Q}_3}={\tilde {\bf Q}^T_3} {\tilde {\bf Q}_3}$, ${\hat {\bf Q}_4}={\tilde {\bf Q}^T_4} {\tilde {\bf Q}_4}$ being two positive semidefinite matrices where ${\tilde {\bf Q}_3}$ and ${\tilde {\bf Q}_4}$ are denoted as
\begin{equation}\label{hatQ3}\nonumber
     {\tilde {\bf Q}_3}=\left [ \begin{array}{cc}
                      \Re\left\{ \left({\bf I}_N \otimes {\bf R}_{p_1} \right)^{\frac{1}{2}}\right\}
                      & -\Im\left\{ \left({\bf I}_N \otimes {\bf R}_{p_1} \right)^{\frac{1}{2}}\right\}  \\
                      \Im\left\{ \left({\bf I}_N \otimes {\bf R}_{p_1} \right)^{\frac{1}{2}}\right\}   &\Re\left\{ \left({\bf I}_N \otimes {\bf R}_{p_1} \right)^{\frac{1}{2}}\right\}
                      \end{array}\right],
\end{equation}
\begin{equation}\label{hatQ4}\nonumber
    {\tilde {\bf Q}_4}=\left [ \begin{array}{cc}
                      \Re\left\{ \left({\bf I}_N \otimes {\bf R}_{p_2} \right)^{\frac{1}{2}}\right\}
                      & -\Im\left\{ \left({\bf I}_N \otimes {\bf R}_{p_2} \right)^{\frac{1}{2}}\right\}  \\
                      \Im\left\{ \left({\bf I}_N \otimes {\bf R}_{p_2} \right)^{\frac{1}{2}}\right\}   &\Re\left\{ \left({\bf I}_N \otimes {\bf R}_{p_2} \right)^{\frac{1}{2}}\right\}
                      \end{array}\right].
\end{equation}
Finally, by combing ${\bf a}_1$ and ${\bf a}_2$ as ${\bf a}=[{\bf a}^T_1,{\bf a}^T_2]^T$, the optimization \eqref{Optimiztion3} has the following form
\begin{equation}\label{App3Optimization}\nonumber
\begin{split}
     &\min_{{\bf a}}~~ {\bf a}^T  {\bf P} {\bf a}- {\bf b}^T {\bf a} + {\rm Tr}\left\{ {\bf R}_{s_{13}}+{\bf R}_{s_{23}}\right\} \\
     s.t. ~~
     &{\bf a}^T  {\bf Q}_1 {\bf a} \leq \tau_1,~
     {\bf a}^T  {\bf Q}_2 {\bf a} \leq \tau_2,~
     {\bf a}^T  {\bf Q}_3 {\bf a} \leq \tau^{'}_r
\end{split}
\end{equation}
where ${\bf P}=\left [ \begin{array}{cc}
                       {\bf P}_2 & {\bf 0}  \\
                      {\bf 0} & {\bf P}_1
                      \end{array}\right]$,
${\bf b}=[2{\bf b}^T_2, 2{\bf b}^T_1]^T$,
${\bf Q}_1=\left [ \begin{array}{cc}
                       {\hat {\bf Q}_1 } & {\bf 0}  \\
                      {\bf 0} & {\bf 0}
                      \end{array}\right]$,
${\bf Q}_2=\left [ \begin{array}{cc}
                       {\bf 0}  & {\bf 0}  \\
                      {\bf 0} & {\hat {\bf Q}_2}
                      \end{array}\right]$ and
${\bf Q}_3=\left [ \begin{array}{cc}
                       {\hat {\bf Q}_3}  & {\bf 0}  \\
                      {\bf 0} & {\hat {\bf Q}_4}
                      \end{array}\right]$.
Since ${\bf P}$ and ${\bf Q}_i$, $i=1,2,3$, are positive semidefinite,
then by definition
the optimization problem \eqref{Optimiztion3} is transformed into the convex QCQP programming problem.

\section{Proof of Lemma 5}
\label{prof_Lemma4}
Due to the
similarity between ${\hat J}_1$ and ${\hat J}_2$, we next focus on deriving the upper bound of ${\hat J}_1$
and the similar results will hold for
${\hat J}_2$.
By defining
${\bf C}=\sigma^2_1 {\bf B}_{g_1}+
{\sigma}^2_r {\bf \Lambda}_{g} {\bf \Lambda}_{A_r}{\bf B}_h {\bf \Lambda}_{A_r}{\bf \Lambda}_{g}$, ${\bf D}={\bf \Lambda}_{g} {\bf \Lambda}_{A_r} {\bf \Lambda}_{h_2} {\bf \Lambda}_{A_2}$,
the MSE in \eqref{eqn:MSEnew1} is rewritten as
\begin{equation} \label{eqn:Append1J1}\nonumber
\begin{split}
    {\hat J}_1 &={\rm Tr}\left \{\left [ {\bf I}_N + {\bf D} {\bf C}^{-1}{\bf D} \right ]^{-1}\right \} \\
    &={\rm Tr}\left [ {\bf I}_N - \left ( {\bf I}_N + {\bf D}^{-1} {\bf C}{\bf D}^{-1}\right )^{-1} \right ],
\end{split}
\end{equation}
where we have used the matrix inversion lemma $\left({\bf I}+{\bf A}^{-1}\right )^{-1}={\bf I}-\left({\bf I}+{\bf A}\right )^{-1}$. Since for any positive definite square matrix ${\bf A}$, it has
${\rm Tr}\left \{ {\bf A}^{-1} \right \}\geq \sum_i \left[ {\bf A}(i,i)\right ]^{-1}$ \cite{RonghongMo2009}, we thus have
\begin{equation} \label{eqn:Append1Inequation2}
\begin{split}
    {\hat J}_{1} &\leq N - \sum^N_{i=1} {\left[\left ( {\bf I}_N + {\bf D}^{-1} {\bf C}{\bf D}^{-1}\right )\left(i,i\right) \right ]^{-1}} \\
    & = {\rm Tr} \left [ {\bf I}_N - \left ( {\bf I}_N + {\bf D}^{-1} {\bf \Lambda}_{C}{\bf D}^{-1}\right )^{-1}\right ] \\
    &={\rm Tr} \left \{ \left [ {\bf I}_N + {\bf D} {\bf \Lambda}^{-1}_{C}{\bf D} \right ]^{-1} \right \}.
\end{split}
\end{equation}
Thus, Lemma~5 is proven.

\section{Deriving the conclusion in \eqref{eqn:KKT8}}
\label{prof_PowerAr}
The Lagrangian function of \eqref{eqn:subprobelem1} is given as
\begin{equation} \label{eqn:Lagrangefunction} \nonumber
\begin{split}
    {\cal L}=& \sum^N_{n=1} \left [\frac{\sigma^2_1 \lambda^n_{Bg_1}+ \sigma^2_r \lambda^n_{Bh} p^n_g p^n_{A_r} }
    {\sigma^2_1 \lambda^n_{Bg_1}+ \sigma^2_r \lambda^n_{Bh} p^n_g p^n_{A_r}+ p^n_{h_2} p^n_g p^n_{A_2}p^n_{A_r}} +
    \frac{\sigma^2_2 \lambda^n_{Bg_2}+ \sigma^2_r \lambda^n_{Bh} p^n_g p^n_{A_r}}
    {\sigma^2_2 \lambda^n_{Bg_2}+ \sigma^2_r \lambda^n_{Bh} p^n_g p^n_{A_r}+ p^n_{h_1} p^n_g p^n_{A_1}p^n_{A_r}} \right] + \\
    & \mu \left[ \sum^N_{n=1} p^n_{A_r} \left( p^n_{h_1} p^n_{A_1} + p^n_{h_2} p^n_{A_2} + \sigma^2_r \lambda^n_{Bh}\right) -\tau_r\right] -
    \sum^N_{n=1} \beta^n p^n_{A_r},
\end{split}
\end{equation}
where $\mu$ and $\beta^n$ are Lagrangian multipliers.
The resultant set of KKT conditions are obtained as
\begin{equation} \label{eqn:KKT1}
\begin{split}
    \frac{\partial {\cal L}}{\partial p^n_{A_r}} =
    & \sum^2_{i=1}\frac{-\sigma^2_i \lambda^n_{Bg_i} p^n_{h_{\bar i}} p^n_g p^n_{A_{\bar i}}}
    {\left[\sigma^2_i \lambda^n_{Bg_i} + p^n_{A_r} \left(\sigma^2_r \lambda^n_{Bh} p^n_g + p^n_g p^n_{h_{\bar i}} p^n_{A_{\bar i}}\right)\right]^2} \\
    & +\mu \left( p^n_{h_1} p^n_{A_1} + p^n_{h_2} p^n_{A_2} + \sigma^2_r \lambda^n_{Bh}\right)
    - \beta^n =0,
\end{split}
\end{equation}
\begin{equation} \label{eqn:KKT2}
    \mu \left[ \sum^N_{n=1} p^n_{A_r} \left( p^n_{h_1} p^n_{A_1} + p^n_{h_2} p^n_{A_2} + \sigma^2_r \lambda^n_{Bh}\right) -\tau_r\right]=0,~
    \beta^n p^n_{A_r} =0, ~\forall n,~
    \mu \geq0,~\beta^n \geq0, ~\forall n
\end{equation}
Based on \eqref{eqn:KKT1} and \eqref{eqn:KKT2}, we have
\begin{equation} \label{eqn:KKT5}
\begin{split}
    & p^n_{A_r}\left( \mu \left( p^n_{h_1} p^n_{A_1} + p^n_{h_2} p^n_{A_2} + \sigma^2_r \lambda^n_{Bh}\right)-
    \sum^2_{i=1}\frac{\sigma^2_i \lambda^n_{Bg_i} p^n_{h_{\bar i}} p^n_g p^n_{A_{\bar i}}}
    {\left[\sigma^2_i \lambda^n_{Bg_i} + p^n_{A_r} \left(\sigma^2_r \lambda^n_{Bh} p^n_g + p^n_g p^n_{h_{\bar i}} p^n_{A_{\bar i}}\right)\right]^2} \right) \\
    &=p^n_{A_r}\beta^n  =0.
\end{split}
\end{equation}
To satisfy \eqref{eqn:KKT5}, we discuss the following cases:

If $\mu \left( p^n_{h_1} p^n_{A_1} + p^n_{h_2} p^n_{A_2} + \sigma^2_r \lambda^n_{Bh}\right) \geq \frac{ p^n_{h_2} p^n_g p^n_{A_2}}{\sigma^2_1 \lambda^n_{Bg_1}} +\frac{ p^n_{h_1} p^n_g p^n_{A_1}}{\sigma^2_2 \lambda^n_{Bg_2}}$,
we must have
    $p^n_{A_r} =0$.

Else, $\mu \left( p^n_{h_1} p^n_{A_1} + p^n_{h_2} p^n_{A_2} + \sigma^2_r \lambda^n_{Bh}\right) < \frac{ p^n_{h_2} p^n_g p^n_{A_2}}{\sigma^2_1 \lambda^n_{Bg_1}} +\frac{ p^n_{h_1} p^n_g p^n_{A_1}}{\sigma^2_2 \lambda^n_{Bg_2}}$, by combining the condition $\beta^n \geq 0$, \eqref{eqn:KKT5} can only be fulfilled with $p^i_{A_r} >0$, This implies the equation \eqref{eqn:add8} given earlier.
Since \eqref{eqn:add8} is a monotonical function of $p^n_{A_r}$ within $(0, +\infty)$, we choose the only positive root of \eqref{eqn:add8} as $p^n_{A_r}$. By combining two cases, we derive the conclusion in \eqref{eqn:KKT8}.

\bibliographystyle{IEEEtran}
\bibliography{IEEEabrv,reference}

\end{document}